\documentclass[%
 reprint,
 amsmath,amssymb,
 aps,
prb,
floatfix,
superscriptaddress,
]{revtex4-2}

\usepackage{graphicx}
\usepackage{mathtools}
\usepackage{xcolor}
\usepackage[dvipsnames]{xcolor}
\usepackage{bm}
\usepackage{cancel}
\usepackage[colorlinks=True,linkcolor=blue,citecolor=blue,urlcolor=blue]{hyperref}

\begin{document}
\preprint{APS/123-QED}

\title{Theory of the Ferroaxial Charge Density Wave in 1T-TiSe$_2$}

\author{Ilaria Maccari}
\affiliation{Institute for Theoretical Physics, ETH Zurich, 8093 Zurich, Switzerland}

\author{Rafael M. Fernandes}
\affiliation{Department of Physics, The Grainger College of Engineering, University of Illinois Urbana-Champaign, Urbana, Illinois 61801, USA}
\affiliation{Anthony J. Leggett Institute for Condensed Matter Theory, The Grainger College of Engineering, University of Illinois Urbana-Champaign, Urbana, Illinois 61801, USA}

\author{Fernando de Juan}
\affiliation{Donostia International Physics Center, 20018 Donostia-San Sebastian, Spain}
\affiliation{IKERBASQUE, Basque Foundation for Science, 48013 Bilbao, Spain}

\author{Maria N. Gastiasoro}
\affiliation{Donostia International Physics Center, 20018 Donostia-San Sebastian, Spain}

\date{\today}

\begin{abstract} Recent experiments using strain to probe the charge-density wave (CDW) of 1T-TiSe$_2$ have uncovered additional transitions and broken symmetries inside the CDW state. The elastocaloric effect reveals a second transition below $T_\mathrm{CDW}$, which elastoresistivity shows to be associated with ferroaxial order, breaking all vertical mirror symmetries while preserving inversion. In addition, at even lower temperatures, a diverging nematic susceptibility has been reported through elastoresistivity. In this work, we argue that this rich phenomenology cannot be explained by the standard $L_1^-$ CDW order parameter, and instead requires the existence of a secondary CDW order parameter. Based on symmetry, microscopic, and diffraction considerations, we propose this subleading CDW to have $L_2^-$ symmetry. By analyzing the Landau theory of coupled $L_1^-$ and $L_2^-$ orders in the presence of strain, we reproduce the phase boundaries observed in the elastocaloric measurements and the emergence of ferroaxial order below the CDW transition. Moreover, the model supports a third threefold-rotational symmetry breaking transition at lower temperatures, consistent with the reported divergence of the nematic susceptibility.
\end{abstract}

\maketitle

\section{Introduction}

Phase transitions in quantum materials can be 
associated with the emergence of different types of ferroic order, such as the familiar ferroelectricity and ferromagnetism. Ferroaxiality is a more elusive type of ferroic order whose ferroic moment is a macroscopic time-reversal invariant axial vector~\cite{Dubovik86,Hlinka16}.
Classically, a non-zero ferroaxial (FA) moment, also known as an electric toroidal dipole or ferro-rotational moment, is generated by a non-irrotational electric polarization field (field with non-vanishing curl), such as a vortex-like distribution of electric dipoles. In a solid, it is an emergent order that can arise from electronic and/or lattice degrees of freedom.   

Being even under time reversal, a ferroaxial moment cannot couple linearly to magnetic fields \cite{Singh25,Cheong2018,ezra2025}. Unlike the ferroelectric moment, the ferroaxial moment is even under both inversion and a reflection with respect to the mirror perpendicular to it, whereas it is odd under reflections with respect to mirrors parallel to it. As such, electric fields cannot couple linearly with the ferroaxial moment either.  Moreover, despite its non-irrotational nature, ferroaxiality is sharply distinct from chirality by the presence of inversion symmetry. In any crystal, a ferroaxial moment necessarily induces electric multipoles of higher angular momentum $l$ due to the breaking of continuous rotation symmetry \cite{Hayami2018}. Focusing on non-cubic crystals (axial groups) that are not themselves ferroaxial, an in-plane FA moment necessarily triggers an electric quadrupole moment ($l=2$). However, an out-of-plane FA moment in these types of crystals only induces $l>2$  multipoles for tetragonal, trigonal, and hexagonal groups. Consequently, strain, which transforms as an electric quadrupole, cannot couple linearly to the FA moment in these cases. Therefore, the absence of a simple field conjugate to ferroaxial moments (electric, magnetic, and strain fields) makes ferroaxial transitions difficult to detect \cite{Hlinka16}. Yet, renewed interest in the subject has recently uncovered several ferroaxial candidate materials and interesting ways to probe and manipulate ferroaxial moments \cite{Jin20,Hayashida20,Hayashida21,Hayami23,Liu23TaS2,Alekseev2024,ezra2025,Zeng25,Singh25,Barantani26,oiwa2026}. Among these candidate materials, recent experiments reported evidence for ferroaxial order in the charge-density wave (CDW) phase of the trigonal transition-metal dichalcogenide 1T-TiSe$_2$ \cite{sarah2026,Ian2026}.

The symmetries broken by the CDW phase transition in 1T-TiSe$_2$ have been the subject of a long debate. While the CDW was initially described in terms of a single $2 \times 2 \times 2$ transition that doubles the lattice periodicity in all three directions (i.e., with wavevectors at the $L$ points of the hexagonal Brillouin zone) and breaks only translation symmetry ~\cite{disalvo1976,holt2001} ($P\bar{3}m1 \rightarrow P\bar{3}c1$), early scanning tunneling microscopy (STM) experiments reported the breaking of both threefold rotational $C_3$ and vertical (dihedral) mirror $\sigma_d$ symmetries, and characterized the CDW as chiral~\cite{ILS10}.
This led to theoretical proposals of a chiral CDW~\cite{vanWezel11,van2012chiral}, which was claimed to emerge at a second transition observed in specific heat measurements~\cite{Castellan13}. Subsequent experiments have both supported~\cite{Iavarone12,Peng22,Xiao24,Kim24seamless,kim2024} and challenged~\cite{Hildebrand18,Lin19,ueda2021,ZhangSHG22,Ueda25} this claim, leading to a continued debate (see Ref. \cite{dani2023} for a more detailed account). It is important to note that, while the initial observations were interpreted in terms of a chiral CDW, several of them are equally consistent with ferroaxial order. This is because some of the probes (e.g., STM) and some of the experimental setups (e.g., optics) explicitly break the inversion symmetry of 1T-TiSe$_2$. In this case, it is not possible to unambiguously distinguish a chiral order parameter from a ferroaxial moment \cite{Kusunose2024}.

Recently, a new set of experiments using strain as a probing field, which leave the inversion symmetry intact, has uncovered additional features below the CDW transition
\cite{sarah2026,Ian2026}. First, elastocaloric effect measurements confirmed the existence of a second transition around $7$ K below $T_\mathrm{CDW}\approx 200$ K.  Second, off-diagonal elastoresistivity, which is enabled by the breaking of all vertical mirror symmetries and hence is a direct probe of ferroaxial order~\cite{ezra2025}, was found to onset right below $T_\mathrm{CDW}$. Third, second-harmonic generation measurements in the same work found no fingerprints of inversion symmetry breaking \cite{Ian2026}. 
Finally, a divergent nematic susceptibility was observed \cite{Lv26,sarah2026,Ian2026} to peak at even lower temperatures, consistent with earlier work \cite{Nunex83}. These combined observations were interpreted as evidence for the emergence of ferroaxial order and, subsequently, of nematic order as the temperature is lowered inside the CDW phase. 

In this work, we employ Landau theory to show that this sequence of broken symmetries cannot be reproduced with the standard $L_1^-$ order parameter (OP) identified in scattering probes at the CDW transition \cite{disalvo1976,holt2001}. This, combined with microscopic considerations, motivates the introduction of a secondary order parameter with $L_2^-$ symmetry. We then show that the coupled $L_1^-+L_2^-$ Landau theory leads to a scenario with a second transition below the CDW transition where FA order with an out-of-plane moment is established, followed by a third transition where nematic order sets in. 

This hierarchy of symmetry-breaking transitions is consistent with experiments, although we caution that the third (nematic) transition is in a range of parameters where the applicability of the Landau theory may be compromised. To test this scenario further, we present calculations of the elastocaloric coefficient under strain and show that all the phase boundaries observed in the experiment are reproduced, further supporting our proposal of a secondary $L_2^-$ CDW and FA order. 

The paper is organized as follows. In Sec.~\ref{sec:L1L2}, we review the standard $L_1^-$ CDW transition and motivate a secondary $L_2^-$ CDW from symmetry, microscopic, and diffraction considerations. In Sec.~\ref{sec:GL}, we derive the Landau theory of the coupled $L_1^-$ and $L_2^-$ order parameters, express it in terms of ferroaxial and nematic composite order parameters, and classify the resulting hierarchies of phase transitions. In Sec.~\ref{sec:strain}, we compute the strain–temperature phase diagram and its thermodynamic signatures, and compare them with experiment. We summarize our results and discuss their broader implications in Sec.~\ref{sec:conclusions}.

\section{The standard $L_1^-$ CDW transition and microscopic motivation for a secondary $L_2^-$ CDW}
\label{sec:L1L2}

\begin{figure}
    \centering
\includegraphics[width=0.9\columnwidth]{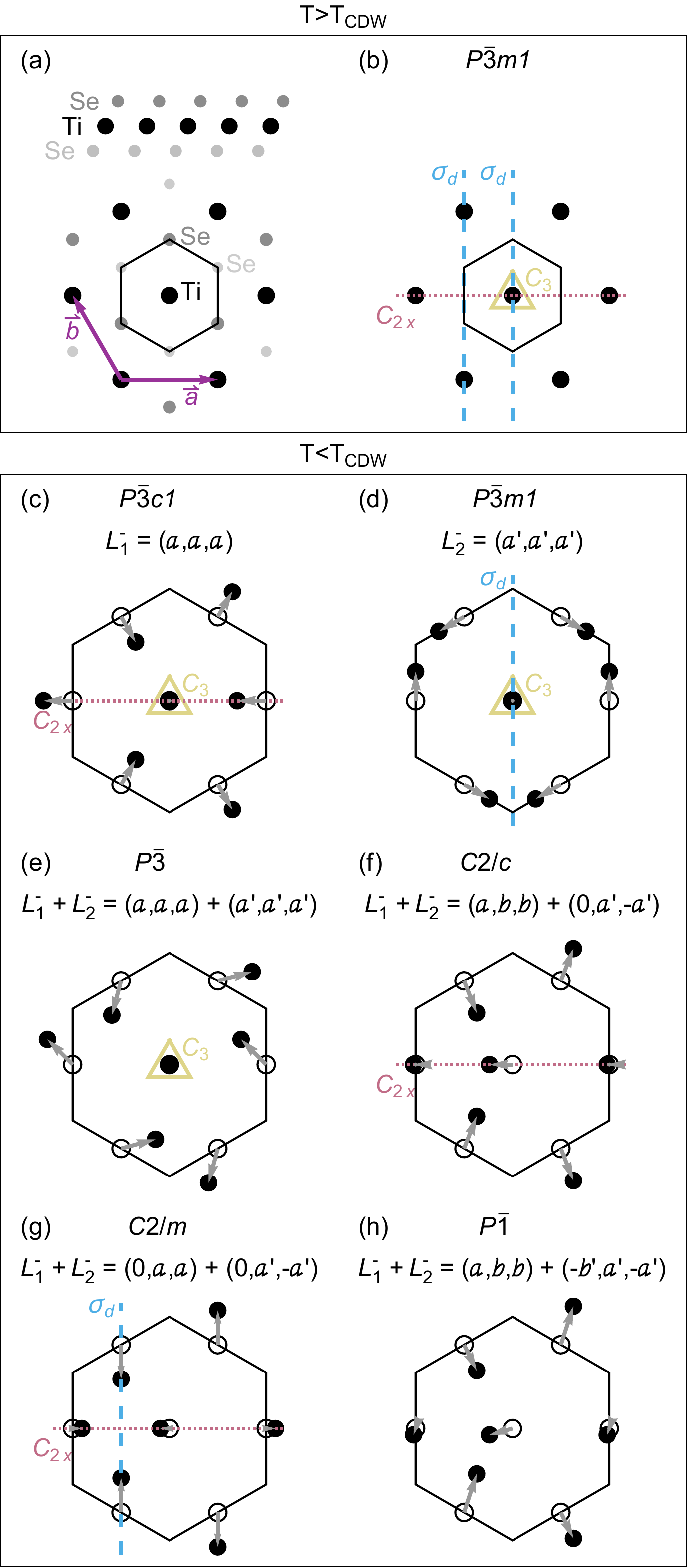}
    \caption{(a) Side and top views of a TiSe$_2$ layer above the CDW transition with its hexagonal unit cell and lattice vectors. (b) Symmetries of the unit cell (vertical mirrors $\sigma_d$, $C_{2x}$ axis, and $C_3$ axis) with the Se atoms omitted for clarity. The SG is listed as well. (c)-(h) In-plane displacement of the Ti atoms on the same layer below $T_\mathrm{CDW}$ (2$\times$2$\times$2 unit cell) for various combinations of the  $L_1^-$ and  $L_2^-$ CDW OPs.  Consecutive layers have opposite displacements. Each panel shows the preserved symmetries of the parent state in (b)  and the SGs. (c) Primary $L_1^-$ OP, (d) secondary $L_2^-$ OP, and (e) FA $\parallel c$ state $L_1^-+L_2^-$. Relevant $C_3$-breaking states are also shown for SGs (f) $C2/c$ (FA $\perp c$), (g)  $C2/m$ (FA  $\perp c$), and (h) $P\bar 1$ (FA  $\parallel c$ and $\perp c$). Note that in (g) the primitive unit cell is a parallelepiped whose base is a 2$\times$1 unit cell, but the hexagonal 2$\times$2 unit cell is shown for ease of comparison.
    }
    \label{fig:uTi}
\end{figure}

\subsection{The $L_1^-$ transition}

1T-TiSe$_2$ crystallizes in a layered trigonal crystal structure of stacked Se-Ti-Se trilayers with each Ti atom octahedrally coordinated by six Se atoms, described by the space group (SG) $P\bar{3}m1$ and illustrated in Fig.~\ref{fig:uTi}(a). Fig.~\ref{fig:uTi}(b) shows some of the symmetry operations of the corresponding point group $D_{3d}$: threefold out-of-plane rotation ($3_{001}$, denoted hereafter $C_3$), twofold in-plane rotation ($2_{100}$, denoted hereafter $C_{2x}$), and vertical mirror reflection ($m_{100}$, denoted hereafter $\sigma_d$). 
Below $T_\mathrm{CDW}$ neutron scattering~\cite{disalvo1976} and x-ray diffraction~\cite{holt2001,ueda2021} measurements report a second-order-like CDW transition at the three commensurate wavevectors $\vec{Q}_n$ ($n=1,2,3$) related by threefold rotations, corresponding to the $L$ points of the hexagonal Brillouin zone, $\vec{Q}_1=(0,1/2,1/2)$, $\vec{Q}_2=(-1/2,0,1/2)$, $\vec{Q}_3=(1/2,-1/2,1/2)$ expressed in terms of the reciprocal lattice vectors $a^*$, $b^*$ and $c^*$ associated with the lattice vectors of Fig.~\ref{fig:uTi}(a). The formation of this $3Q$ CDW results in a doubling of the hexagonal lattice parameters and an enlarged $2\times2\times2$ unit cell. Consequently, the Ti atoms are displaced from their original crystallographic positions in Fig.~\ref{fig:uTi}(a) to the new positions illustrated in Fig.~\ref{fig:uTi}(c). 

Neutron data~\cite{disalvo1976} found no intensity in peaks in the $(h0l)$ plane, which was interpreted as a glide extinction from $m_{010}\{0,0,\tfrac{1}{2}\}$: if the displacements are even under the glide, then Bragg peaks with half integer $l$ should be extinct. This in turn indicates the atomic displacements to be purely transverse. From the four possible irreducible representations (irreps) at the $L$ point, $L_{1/2}^\pm$ (see Table~\ref{tab:c2h}), only $L_1^-$ and $L_2^+$ involve transverse displacements, and structure refinement concluded that the data was consistent with $L_1^-$, shown in Fig.~\ref{fig:uTi}(c). Since the star of $L$ contains three wavevectors, the corresponding space-group irrep is three-dimensional, and the $3Q$ CDW corresponds to the configuration $\vec{\psi}=(a,a,a)$ of the $L_1^-$ order parameter, where $a$ is proportional to the amplitude of the transverse displacements.

\begin{table}[]
    \centering
    \begin{tabular}{c|c c c c}
         $C_{2h}$ & $E$ & $C_{2x}$ & $\sigma_d$ & $i$ \\
         \hline
         $L_1^+$ & 1 & 1 & 1 & 1 \\
         $L_1^-$ & 1 & 1 & -1 & -1 \\
         $L_2^+$ & 1 & -1 & -1 & 1 \\
         $L_2^-$ & 1 & -1 & 1 & -1
    \end{tabular}
    \caption{Character table for the little group at the $L$ point, which is isomorphic to $C_{2h}$. The symmetry operations $C_{2x}$ (in-plane twofold rotation) and $\sigma_d$ (vertical mirror reflection) are shown in Fig.\ref{fig:uTi}, whereas $i$ is inversion. }
    \label{tab:c2h}
\end{table}

\subsection{Ferroaxial order}

Evidence for ferroaxial order was obtained from elastoresistivity measurements in Refs. \cite{sarah2026,Ian2026}. These experiments signaled the breaking of the vertical mirror symmetry at a temperature $T_\mathrm{FA}$ below $T_\mathrm{CDW}$, but well above the divergence of the nematic susceptibility at $T_{\rm nem}$, implying that $C_3$ symmetry is preserved. This reduces the point-group symmetry from $D_{3d}$ to $C_{3i}$ (also denoted $S_6$) and allows the existence of a FA moment along the $c$ axis, which transforms as the $A_{2g}$ irrep of $D_{3d}$. At temperatures below $T_{\rm nem}$, the $C_3$ symmetry is additionally broken by a nematic (or quadrupolar) order parameter, further lowering the point group from $C_{3i}$ to $C_{i}$ (equivalently, from $S_6$ to $S_2$). In $D_{3d}$, the nematic order parameter can also be characterized as an in-plane FA moment, since both transform as the $E_g$ irrep. This is a consequence of the fact that, in $D_{3d}$, the nematic and planar FA orders are indistinguishable.

While nematicity (or in-plane ferroaxiality) can emerge from $L_1^-$ alone \cite{dani2023}, such as in the case of a 1Q CDW, the emergence of an out-of-plane FA moment while preserving $C_3$ symmetry is impossible with the condensation of $L_1^-$ alone. To explain the elastoresistivity experiments, a secondary mode is therefore required. Since an out-of-plane FA moment transforms as the $A_{2g}$ point-group irrep, it can arise from the coexistence of $L_1^-$ with another CDW order parameter that transforms as $L_2^-$, $M_2^+$ or $A_2^-$ \cite{sarah2026}. Here, we consider the case in which the additional (but sub-leading) CDW instability has $L_2^-$ character. As we will see, this choice is motivated by a combination of symmetry analysis, DFT calculations (reported elsewhere), a $\mathbf{k}\cdot \mathbf{p}$ model of the electronic energy gain from the CDW distortion~\cite{dani2023}, and recent X-ray data.  

\subsection{Proposal for $L_2^-$}

Among the three proposed candidate modes, $L_2^-$, $M_2^+$ or $A_2^-$, it can be argued that $L_2^-$ is the most energetically competitive. First-principles calculations find that the energies of the $L_2^-$ and  $A_2^-$ phonon modes are below the energy scale associated with the CDW transition temperature~\cite{Bianco15,Hellgren17,KimUnpublished}, and the condensation energy difference between CDW states with $L_1^-$ and $L_2^-$ symmetries differs only by 0.5 meV per formula unit \cite{Wegner2020}.  The reason why these two modes have similar condensation energies can be explained by examining the low energy band structure of TiSe$_2$ near the Fermi level, which is composed of a doublet valence band around $\Gamma$ 
with $\Gamma_3^-$ symmetry, derived from Se $(p_x,p_y)$ orbitals, and a conduction band pocket around each of the $L$ points with symmetry $L_1^+$, derived from Ti $3d$ orbitals \cite{Zunger78}. A CDW distortion with momentum $L$ folds these states back to $\Gamma$ and can thus open a gap at the crossing points between folded conduction and valence bands if the symmetry of the CDW order parameter matches the product irrep of such bands. Since $\Gamma_3^- \otimes L_1^+ = L_1^-+L_2^-$, a CDW distortion with either symmetry can cause an electronic energy gain by opening a gap between the folded and unfolded bands. This argument singles out $L_2^-$ over $M_2^+$ and $A_2^-$ as the most likely candidate for a secondary mode. In fact, in a simplified model where both conduction and valence bands have full rotational symmetry, the electronic energy gain from $L_1^-$ and $L_2^-$ is exactly the same. Including the known anisotropy of the bands results in larger energy gain for $L_1^-$ than $L_2^-$~\cite{dani2023}.  

While describing the full CDW instability in TiSe$_2$ is a more complex problem that requires knowledge of the lattice energetics beyond the harmonic approximation as well as electronic interactions, symmetry analysis combined with previous DFT results makes it plausible that while the leading instability is in the $L_1^-$ channel, a secondary one with $L_2^-$ symmetry may also lead to further electronic energy gain. A further argument that supports such a mode can be made by considering the role of the additional valence band pocket with symmetry $\Gamma_2^-$~\cite{Zunger78},  which is derived from the $p_z$ orbitals of Se. This pocket was not believed to be relevant for the CDW transition because it is only coupled to conduction modes by an $L_1^-$ mode through spin-orbit coupling, which is a subleading effect. However, since $\Gamma_2^- \otimes L_1^+ = L_2^-$, this pocket may lead to further condensation energy gain in the presence of the secondary $L_2^-$ distortion. Another motivation to consider $L_2^-$ is that the glide extinction in the $(h0l)$ plane in the original neutron data~\cite{disalvo1976} has been revisited.  Recent x-ray scattering measurements report a weak $(1/2,0,9/2)$ superlattice peak developing at $T \lesssim T_\mathrm{CDW}$ (see the Supplemental Material of Ref.~\cite{kim2024}), although the available data do not resolve whether its onset coincides with $T_\mathrm{CDW}$ or occurs slightly below it. Such a reflection is not allowed by $L_1^-$ order alone and requires breaking the glide symmetry. Thus it is compatible with a combined $L_1^- + L_2^-$ order [Fig.~\ref{fig:uTi}(e)].

An $L_2^-$ mode differs from $L_1^-$ in its $\sigma_d$ and $C_{2x}$ parities: $L_2^-$ is $\sigma_d$ even and $C_{2x}$ odd, while $L_1^-$ is $\sigma_d$ odd and $C_{2x}$ even. Both modes $L_{1/2}^-$ are odd under inversion $i$. $L_1^-$ is thus even under the glide $m_{100}\{0,0,\tfrac{1}{2}\}$, while $L_2^-$ is odd. With the condensation of the secondary $L_2^-$ mode, the system can develop a finite ferroaxial moment along the $c$ axis at $T_\mathrm{FA}<T_\mathrm{CDW}$ because both the glide and $C_{2x}$ are broken, reducing the point group from $D_{3d}$ to $C_{3i}$. Even though both modes are odd under the inversion center of the high temperature structure, located in the Ti site, both modes are even under inversion combined with the $c$ translation, which represents an inversion center located between Ti layers. The combined state with $L_{1}^-$ and $L_{2}^-$ modes always preserves this inversion center. Therefore none of the structures obtained in this work are chiral.  

\begin{figure}
    \centering
\includegraphics[width=0.8\columnwidth]{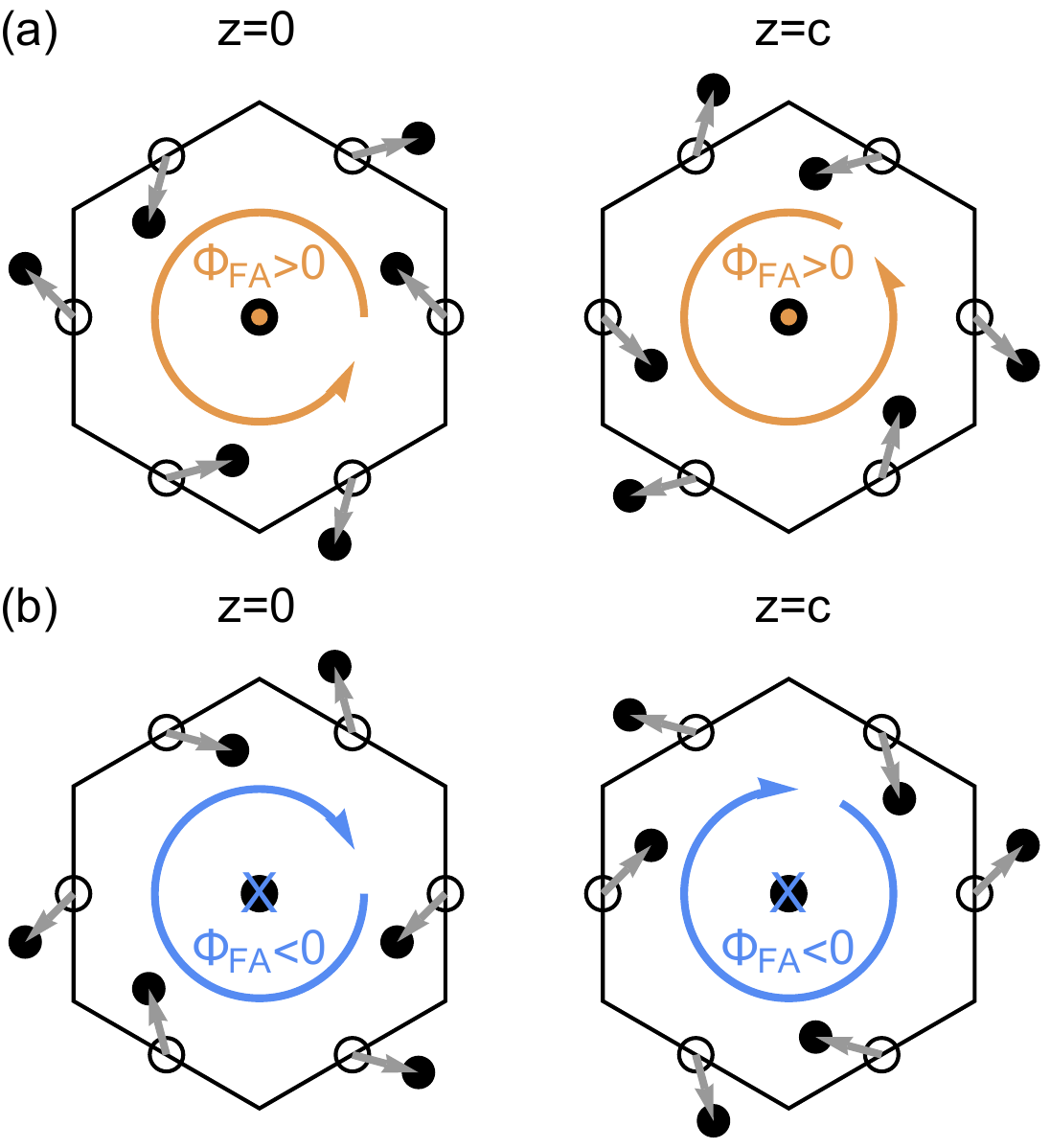}
    \caption{(a) Displacement of the Ti atoms in the FA CDW state $L_1^-+L_2^-=(a,a,a)+(a',a',a')$ for the two consecutive layers $z=0$ and $z=c$, with $\Phi_\mathrm{FA}=\vec\psi\cdot\vec\eta=3aa'>0$ [see also Fig.~\ref{fig:uTi}(e)]. The circular arrows indicate the common sense in which the Ti displacement directions are rotated away from the purely transverse $L_1^-$ pattern  of Fig.~\ref{fig:uTi}(c). This sense, set by the sign of $\Phi_\mathrm{FA}$, is the same in both layers, although the displacements themselves are reversed. 
    (b) The degenerate state $L_1^-+L_2^-=(a,a,a)+(-a',-a',-a')$ with opposite FA moment $\Phi_\mathrm{FA}=-3aa'$.
    }
    \label{fig:FA-moment}
\end{figure}

While the displacements of the $L_1^-$ mode are known, the $L_2^-$ mode allows both in-plane longitudinal and out-of-plane displacements. For simplicity we represent it through in-plane Ti displacements, but our phenomenological theory does not rely on any particular choice, as it is based only on symmetry. The Ti atomic displacements and the space group resulting from the individual condensation of $L_1^-$ and $L_2^-$ are shown in Fig.~\ref{fig:uTi}(c) and Fig.~\ref{fig:uTi}(d), respectively, while the distinct possible configurations arising from the simultaneous condensation of both modes are shown in Fig.~\ref{fig:uTi}(e)-(h). Among them, only panels (e) and (h) correspond to configurations that give rise to a finite ferroaxial moment along the $c$ axis, with the latter additionally breaking rotational $C_3$ symmetry. In these cases, the displacement pattern has a clear non-irrotational character, which further highlights its ferroaxial symmetry. Note that the
FA moment is the same in two consecutive layers $z=0$ and $z=c$, as illustrated in Fig.~\ref{fig:FA-moment} for the two $C_3$-symmetric degenerate states with opposite FA moments.

\section{Landau theory of the coupled CDW states}
\label{sec:GL}

Motivated by the previous considerations, we propose that the CDW transition at high temperature $T_\mathrm{CDW}$ is driven by the condensation of the $L_1^-$ mode, followed at a lower temperature by the condensation of the $L_2^-$ mode.
We now derive the Landau theory for the two coupled three-component CDW order parameters: $\vec{\psi} = (\psi_1, \psi_2, \psi_3)$, associated with the $L_1^-$ mode, and $\vec{\eta} = (\eta_1, \eta_2, \eta_3)$, associated with $L_2^-$. In both cases, each component $n=\{1,2,3\}$ corresponds to one of the three CDW wavevectors $\vec{Q}_n$.
We use the INVARIANTS tool \cite{invariants2003} to obtain the Landau free energy $\mathcal{F}=\mathcal{F}_\psi+\mathcal{F}_\eta+\mathcal{F}_{\psi\eta}$ up to quartic order in the order parameters. For the individual CDW terms, we obtain 
\begin{align}
    \label{eq:FL1}\mathcal{F}_\psi&=a_\psi(T-T_\mathrm{CDW})  |\vec\psi|^2 +  b_\psi|\vec\psi|^4   +c_\psi \left(\psi_1^4+\psi_2^4+\psi_3^4\right) \\
\mathcal{F}_\eta&=a_\eta(T-T_\eta)|\vec\eta|^2 +  b_\eta |\vec\eta|^4 +c_\eta \left(\eta_1^4+\eta_2^4+\eta_3^4\right)\label{eq:FL2}
\end{align}
where $|\vec\psi|^2=\sum_n  \psi_n^2$, and similarly for $\vec\eta$.
Here, the quadratic coefficients satisfy  $a_\psi,a_\eta >0$ while the quartic coefficients satisfy $b>0$ and $b+c>0$ to ensure that the free energy is bounded from below. For both $\mathcal{F}_\psi$ and $\mathcal{F}_\eta$, when the quadratic term becomes negative, the system condenses into a state with a finite order parameter. The structure of this state is selected by the sign of $c$: for $c<0$, the order parameter takes the form of a $1Q$ state $(a,0,0)$, whereas for $c>0$ it selects the $C_3$-symmetric $3Q$ state $(a,a,a)$. Since the latter is the experimentally observed state below the CDW transition (the nematic susceptibility peaks around 28 K below the second transition \cite{sarah2026}), we set $c_\psi>0$ and $c_\eta>0$, and the uncoupled critical temperatures $T_\mathrm{CDW}>T_\eta$ appearing in the quadratic terms. 

The lowest order coupling between $\vec\psi$ and $\vec \eta$ has the following quartic terms,
\begin{align}
    \label{eq:FL1L2}\mathcal{F}_{\psi\eta}&= d_1 (\vec\psi\cdot\vec \eta)^2 +d_2\sum_{n}\psi_n\eta_n \psi_{n+1}\eta_{n+1} +d_3|\vec\psi|^2|\vec \eta|^2 \\\nonumber
&+\sum_n\psi_n\eta_n \left[ d_4\left(\psi_{n+2}^2 - \psi_{n+1}^2\right) + d_5\left(\eta_{n+2}^2 - \eta_{n+1}^2\right)\right]
\end{align}
where the sums run over $n=1,2,3$, with indices defined modulo 3. 

To better understand the possible sequences of phase transitions described by Eqs. \eqref{eq:FL1}-\eqref{eq:FL1L2}, it is instructive to decompose the bilinear of the order parameters $\vec{\psi}$ and $\vec{\eta}$ into combinations that have $C_3$-symmetric, ferroaxial or nematic characters. For a generic field $\vec{v} \in \{\vec{\psi}, \vec{\eta}\}$, we define the two-component nematic order parameter, which transforms as the $E_{g}$ irrep of the point group $D_{3d}$ as:
\begin{equation}
    \label{eq:nem_v}
    \vec{N}_v \equiv \left(v_1^2 - \frac{1}{2}(v_2^2+v_3^2), \frac{\sqrt{3}}{2}(v_2^2 - v_3^2) \right).
\end{equation}
The free energy of the uncoupled order parameters from Eqs. \eqref{eq:FL1}-\eqref{eq:FL2} can then be  rewritten as:
\begin{equation}
    \mathcal{F}_{v=\psi, \eta} = a_v (T-T_v)|\vec{v}|^2 + b_v |\vec{v}|^4 + \frac{c_v}{3} \left(|\vec{v}|^4 + 2|\vec{N}_v|^2 \right)
    \label{eq:Fv_nem}
\end{equation}
To rewrite the coupling free energy in Eq.~\eqref{eq:FL1L2} in the same spirit, we introduce the ferroaxial order parameter parallel to the $c$ axis, which transforms as the $A_{2g}$ irrep of $D_{3d}$: 
\begin{equation}
    \Phi_\mathrm{FA}\equiv\vec{\psi} \cdot \vec{\eta},
    \label{eq:FA}
\end{equation} 
and another $E_g$ mixed nematic order parameter in $\vec{\psi}$ and $\vec{\eta}$:
\begin{equation}
    \vec{N}_{\psi \eta}\equiv\left( \frac{\sqrt{3}}{2}(\psi_2\eta_2-\psi_3\eta_3),-\psi_1\eta_1 + \frac{1}{2}(\psi_2\eta_2+\psi_3\eta_3) \right).
    \label{eq:compositenem}
\end{equation}
Note that, since $\vec{\psi}(L_1^-)$ and $\vec{\eta}(L_2^-)$ transform with opposite signs under $C_{2x}$ (Table \ref{tab:c2h}), the components of $\vec{N}_{\psi\eta}$ are rotated relative to those of $\vec{N}_v$ in Eq.~\eqref{eq:nem_v}.
Using Eqs. \eqref{eq:nem_v},\eqref{eq:FA} and \eqref{eq:compositenem} we can express the cross-coupling free energy $\mathcal{F}_{\psi\eta}$ as:
\begin{equation}
\begin{split}
    \mathcal{F}_{\psi\eta}=&\left(d_1+\frac{d_2}{2} \right)\Phi_\mathrm{FA}^2 + \left(d_3 - \frac{d_2}{6}\right)|\vec{\psi}|^2|\vec{\eta}|^2\\ & - \frac{d_2}{3}\vec{N}_\psi\cdot\vec{N}_\eta   +\frac{2}{\sqrt{3}} \left( d_4 \vec{N}_\psi + d_5  \vec{N}_\eta\right) \cdot \vec{N}_{\psi \eta}.  
\end{split}
\label{eq:FL1L2v2}
\end{equation}

This form makes it transparent which specific states each coupling coefficient $d_i$ favors.
Both $d_1 < 0$ and $d_2 < 0$ promote FA order ($\Phi_\mathrm{FA}\neq 0$), which is $C_3$ symmetric when $c_\eta, c_\psi$ are positive and large enough. Conversely, a repulsive $d_1 > 0$ breaks $C_3$ and  suppresses $\Phi_\mathrm{FA}$; however, when the quartic terms in  Eqs. \eqref{eq:FL1}-\eqref{eq:FL1L2} favor two 3Q states, $\vec{\psi}$ and $\vec{\eta}$ cannot be made orthogonal, such that $\Phi_{\mathrm{FA}}$ remains finite. A sufficiently large $d_2 > 0$ instead favors nematic order at the expense of FA moment along the $c$ axis, whereas for a smaller $d_2 > 0$, a coexistence phase with finite nematic vectors, $\vec{N}_\psi$  and $\vec{N}_\eta$, as well as $\Phi_\mathrm{FA}$ is stabilized. 

Since the coupling $d_3$ depends only on the amplitudes $|\vec\psi|$ and $|\vec\eta|$, on its own it leaves the $C_3$ symmetric FA state degenerate with a $C_3$-breaking state differing only in the relative sign of one component $\psi_n\eta_n$ (a degeneracy lifted by $d_1$ or $d_2$).
Its main effect is to renormalize the critical temperature of the secondary order parameter, shifting it closer to $T_\mathrm{CDW}$ for $d_3 < 0$, or pushing it further away for $d_3 > 0$. However, if the biquadratic coefficient becomes large enough, the competition between the two order parameters may not allow for a state where the two CDW order parameters coexist. Finally, $d_4$ and $d_5$, regardless of their sign, stabilize the coexistence of nematic, $\vec{N}_{\psi},  \vec{N}_{\eta}$, mixed nematic, $ \vec{N}_{\psi \eta}$, and FA, $\Phi_\mathrm{FA}$, orders below a certain threshold, above which only the nematic and mixed nematic orders survive. 
In summary, the coupling parameters $d_i$ select among the low-temperature states outlined in Table~\ref{tab:transitions}, along with their resulting space groups, visually represented in panels (e)-(h) of Fig.~\ref{fig:uTi}. For a detailed description of the possible states that each parameter $d_i$ can induce on its own, for a representative set of parameters, see Appendix~\ref{app:di}. 

\begin{table}[]
    \centering
    \begin{tabular}{c|c|c|c|c}
          $(\psi_1,\psi_2,\psi_3)$ & $(\eta_1,\eta_2,\eta_3)$ &  FA $\parallel c$ & Nematic, FA $\perp c$ & Space group \\\hline
          $(a,a,a)$ & $(a',a',a')$ & yes & no & {\color{RoyalPurple}$P\bar 3$}  \\\hline
          $(a,a,b)$ & $(a',-a',0)$ & no  & yes & {\color{PineGreen}$C2/c$} \\\hline
          $(a,a,0)$ & $(-a',a',-b')$ & no  & yes & {\color{LimeGreen}$C2/m$} \\\hline
          $(a,b,c)$ & $(a',b',c')$ & yes  & yes & {\color{magenta}$P\bar 1$} \\\hline
    \end{tabular}
    \caption{Possible coexistence states of $\vec{\psi}$ and $\vec{\eta}$ induced below the $3Q$ CDW transition after the secondary order parameter $\vec{\eta}$ condenses, depending on the $d_i$ coupling parameters in Eq. \eqref{eq:FL1L2v2} (see text). The space group is changed from P$\bar{3}$c1 ($\#165$) (for $\vec{\psi}\neq 0$) to the subgroup specified in the last column: {\color{RoyalPurple}$P\bar 3$} ($\# 147$), {\color{PineGreen}$C2/c$} ($\#15$), {\color{LimeGreen}$C2/m$} ($\#12$) 
    and {\color{magenta}$P\bar 1$} ($\#2$). }
    \label{tab:transitions}
\end{table}

\begin{figure*}
    \centering
    \includegraphics[width=\linewidth]{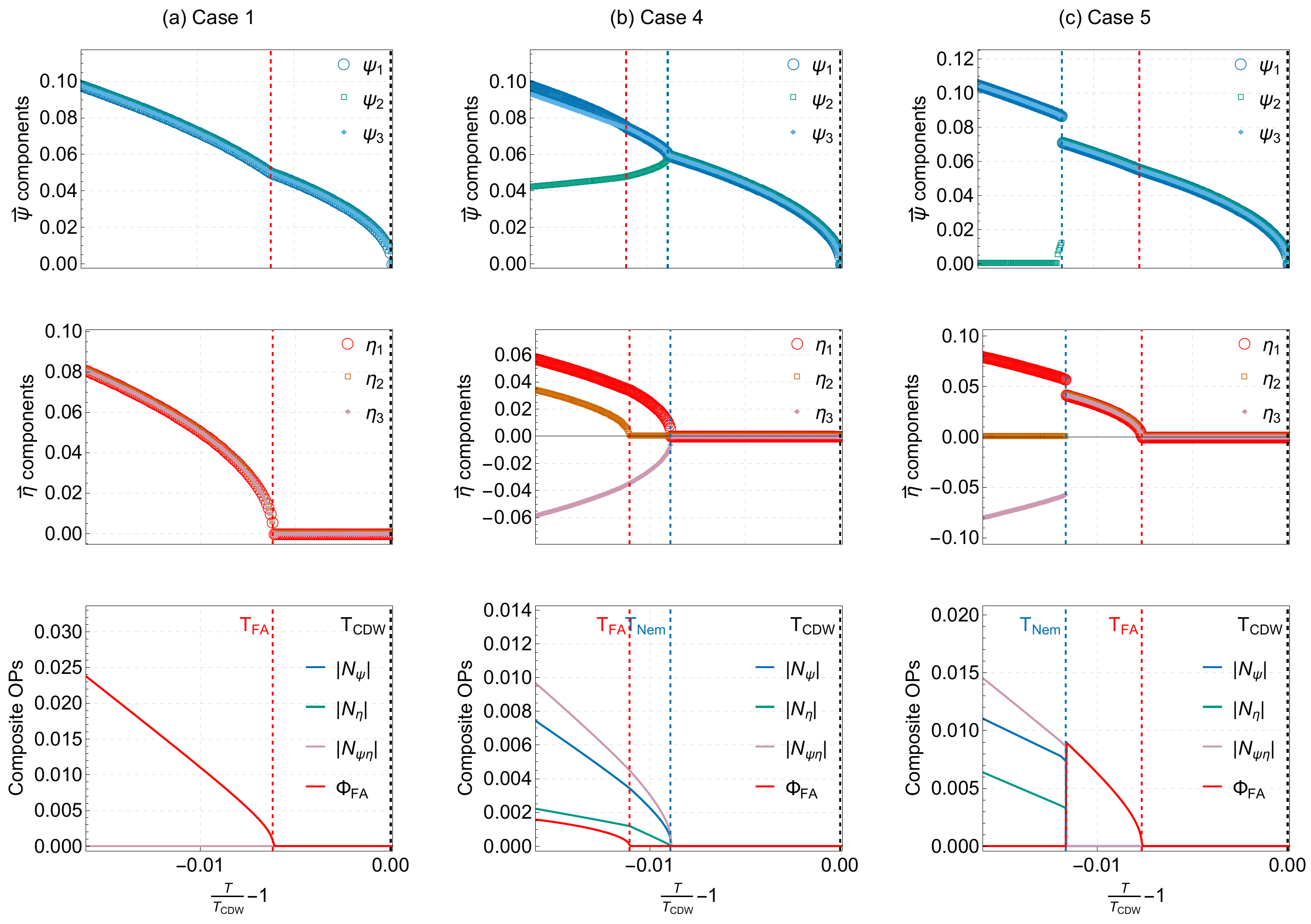}
    \caption{Examples of phase transition hierarchies obtained from minimization of $\mathcal{F}_{\psi}+\mathcal{F}_{\eta}+\mathcal{F}_{\psi\eta}$ (Eqs.~\eqref{eq:Fv_nem} and \eqref{eq:FL1L2v2}). The rows show the primary CDW OP $\vec{\psi}$ (top), the secondary CDW OP $\vec{\eta}$ (middle), and the symmetry-breaking composite OPs $|N_\psi|, |N_\eta|, |N_{\psi \eta}|, \Phi_\mathrm{FA}$ (bottom). In all panels, $T_\mathrm{CDW}=1$, $T_\eta=0.99$, $a_\psi=1$, $a_\eta=0.8$, $2b_{\psi}=2b_\eta= c_\psi=c_\eta=0.5$, but the values of the $d_i$ coefficients are different. Unless otherwise stated, $d_i$ is set to zero. (a) $d_1=-0.4$ (Case 1), $T_\mathrm{CDW}>T_\mathrm{FA}$. Upon cooling, the OP structure evolves as $(a,a,a) + (0,0,0) \to (a,a,a) + (a',a',a')$, with $P\bar{3}c1 \to P\bar{3}$. 
    (b) $d_4=0.4$ (Case 4), $T_\mathrm{CDW}>T_\mathrm{nem}>T_\mathrm{FA}$. The $\vec{\psi}+\vec{\eta}$ OP evolution as temperature is lowered follows $(a,a,a) + (0,0,0) \to (a,a,b) + (a',-a',0) \to (a,b,c) + (a', -b',c')$, with $P\bar{3}c1 \to C2/c \to P\bar{1}$. As a consequence, the three nematic vectors evolve from vanishing in the $P\bar{3}c1$ phase to a collinear configuration in the $C2/c$ phase, where $\vec{N}_\psi$ and $\vec{N}_\eta$ are parallel and $\vec{N}_{\psi\eta}$ is antiparallel to both, and finally to a generically noncollinear configuration in the $P\bar{1}$ phase, accompanied by the emergence of a finite ferroaxial moment $\Phi_\mathrm{FA}$.
    (c) $d_1=-0.2$ and $d_4=d_5=0.4$ (Case 5), $T_\mathrm{CDW}>T_\mathrm{FA}>T_\mathrm{nem}$. The OP evolution as temperature is lowered follows $(a,a,a) + (0,0,0) \to (a,a,a) + (a',a',a') \to (a, 0, a) + (a', 0, -a')$, with $P\bar{3}c1 \to P\bar{3} \to C2/m$. The three nematic vectors below $T_\mathrm{nem}$ are collinear with $\vec{N}_\psi$ and $\vec{N}_\eta$ are parallel and $\vec{N}_{\psi\eta}$ is antiparallel to both.}
    \label{fig:secIII}
\end{figure*}

We now consider the general case where all $d_i$ can be nonzero,  keeping $T_\eta<T_\mathrm{CDW}$ so that the 3Q CDW condenses first (see Appendix \ref{app:numerics} for the explored parameter range and details on the numerical minimization). The following hierarchies of transitions emerge below $T_{\rm CDW}$, with representative examples shown in Fig.~\ref{fig:secIII}:

\begin{enumerate}
\item{ $T_\mathrm{CDW}>T_\mathrm{FA}$ [Fig.~\ref{fig:secIII}(a)], with only FA order ($P\bar{3}$) at low temperatures.}
\item{ $T_\mathrm{CDW}>T_\mathrm{nem}$, with only nematic order at low temperatures. The nematic phase is either $C2/c$, with $\vec\psi\propto(a,a,b)$ and $\vec\eta\propto(a',-a',0)$, or $C2/m$, with $\vec\psi\propto(a,a,0)$ and $\vec\eta\propto(a',-a',0)$, i.e. a 2Q configuration in which both order parameters have no component along the same ordering wave vector. On further cooling, the nematic order can evolve between these configurations, $C2/c\to C2/m$ (with a four-fold cell) $\to C2/m$, the latter with $\vec\psi\propto(a,a,0)$ and $\vec\eta\propto(a',-a',b')$, without the appearance of FA order.}
\item{ $T_\mathrm{CDW}>T_\mathrm{FA}=T_\mathrm{nem}$, with nematic and FA orders appearing simultaneously ($P\bar{1}$). In part of the parameter space, the FA order vanishes again at a lower temperature, leaving a nematic $C2/m$ phase.}
\item{ $T_\mathrm{CDW}>T_\mathrm{nem}>T_\mathrm{FA}$, with two possible realizations upon further cooling:
\emph{(i)} The nematic order persists below the onset of ferroaxial order at $T_\mathrm{FA}<T_\mathrm{nem}$ [Fig.~\ref{fig:secIII}(b)], leading to a $C2/c\to P\bar{1}$ transition. In part of the parameter space, the ferroaxial order disappears again at lower temperatures, leaving a nematic $C2/m$ phase at the lowest temperatures.
\emph{(ii)} The nematic order exists only within an intermediate temperature window and disappears at $T_\mathrm{FA}<T_\mathrm{nem}$, where the system enters a purely ferroaxial $P\bar{3}$ phase, with $\vec\eta$ adopting a 3Q configuration parallel to $\vec\psi$.
}
\item{ $T_\mathrm{CDW}>T_\mathrm{FA}>T_\mathrm{nem}$ [Fig.~\ref{fig:secIII}(c)], with the FA order parameter disappearing at $T_\mathrm{nem}$ concomitantly with the emergence of the nematic order ($P\bar{3}\to C2/c$ or $C2/m$). At $T<T_\mathrm{nem}$, the FA order can re-emerge as $\vec\eta$ evolves from a 2Q to a 3Q configuration, leading to the $P\bar{1}$ phase with both nematic and FA orders.}
\end{enumerate}

Within the explored parameter range, Cases 1 to 4(i) can each be realized with a single nonzero $d_i$, see Table \ref{tab:di}, whereas Cases 4(ii) and 5 require at least two simultaneously nonzero coupling parameters. Importantly, the results of Refs. \cite{sarah2026,Ian2026} are consistent with Case 5. 

The order of the phase transitions can be constrained by symmetry considerations. Within Landau theory, a necessary condition for a continuous transition between phases with space groups $G$ and $H$ is that $H$ be a subgroup of $G$. If this condition is not satisfied, a direct transition between the two phases must be discontinuous. However, the existence of a group--subgroup relation does not guarantee a continuous transition, since the Landau expansion may contain cubic invariants of the critical order parameter, which generically favor a first-order transition at the mean-field level, or have a negative quartic Landau coefficient due to microscopic considerations. 

This distinction is particularly relevant to transitions out of the $P\bar{3}$ ferroaxial phase, where both CDW order parameters are already finite. The onset of nematicity corresponds to a distortion of the existing $C_3$-symmetric CDW configuration. 
As shown explicitly in Appendix~\ref{app:Nem1st}, expanding the quartic Landau functional around this ordered state generates cubic terms in the nematic fluctuations, consistent with the underlying $C_{3i}$ symmetry. These terms generically drive the nematic transition first order at the mean-field level, making the $P\bar{3}\to P\bar{1}$ transition discontinuous. In contrast, a direct $P\bar{3}\to C2/c$ or $P\bar{3}\to C2/m$ transition is necessarily discontinuous because the two space groups do not satisfy a group--subgroup relation.

\section{Phase diagram and thermodynamic features under strain}
\label{sec:strain}

The recent elastocaloric and elastoresistivity experiments reporting ferroaxial order use strain as a probing field. Here, we incorporate strain as an external field into our model and, when possible, compare the resulting temperature-strain phase diagram with thermodynamic data. The strain tensor is formally given in terms of the lattice displacement vector $\vec{u}$ by $\varepsilon_{ij}=\frac{1}{2}(\partial_i u_j + \partial_j u_i)$. 
The three independent in-plane components decompose into an $A_{1g}$ component $\varepsilon_A = \varepsilon_{xx} + \varepsilon_{yy}$ and an $E_g$ component $\varepsilon_E = (\varepsilon_{xx} - \varepsilon_{yy},\, -2\varepsilon_{xy})\equiv (\varepsilon_{E1},\varepsilon_{E2})$, corresponding to the irreps of $D_{3d}$. They generate, to leading order, the following additional terms in the free energy:
\begin{align}
\label{eq:Feps}
\mathcal{F}_\varepsilon=&\left(\lambda_{E\psi} N_{\psi,1}+\lambda_{E\eta}N_{\eta,1}+\lambda_{E\psi\eta}  N_{\psi \eta,1}\right)\varepsilon_{E1}\\\nonumber
+&\left(\lambda_{E\psi}N_{\psi,2}+\lambda_{E\eta}N_{\eta,2}+\lambda_{E\psi\eta} N_{\psi \eta,2}\right)\varepsilon_{E2}\\\nonumber
+&\left(\lambda_{A\psi} |\vec{\psi}|^2+\lambda_{A\eta} |\vec{\eta}|^2\right)\varepsilon_A.
\end{align}

Under adiabatic conditions, the elastocaloric effect (ECE) is expressed in terms of derivatives of the entropy with respect to strain and temperature \cite{Ikeda2019,Ikeda2021}:
\begin{equation}
    \left(\frac{\partial T}{\partial \varepsilon}\right)_S=-\frac{\left(\frac{\partial S}{\partial \varepsilon}\right)_T}{\left(\frac{\partial S}{\partial T}\right)_\varepsilon}=-\frac{T}{C_\varepsilon}\left(\frac{\partial S}{\partial \varepsilon}\right)_T\label{eq:ece}
\end{equation}
where $C_\varepsilon$ is the specific heat under constant strain. Within our Landau model, the entropy is given by 
\begin{equation}
    S=-\partial \mathcal F/\partial T=-a_\psi|\vec{\psi}|^2-a_\eta|\vec{\eta}|^2
\end{equation} 
Hence, by minimizing the free energy in Eqs.~\eqref{eq:FL1}-\eqref{eq:FL1L2} and Eq. \eqref{eq:Feps}, we obtain the order parameter solutions as a function of strain and temperature, $\vec{\psi}(T,\varepsilon)$ and $\vec{\eta}(T,\varepsilon)$. From them, we compute the specific heat  $C_\varepsilon$ and the ECE coefficient in Eq. \eqref{eq:ece}.

For the zero strain free energy in Eqs.~\eqref{eq:FL1}-\eqref{eq:FL1L2}, we choose a parameter regime that corresponds to Case 1, with $T_\mathrm{CDW}>T_\mathrm{FA}$ and no nematic transition. We make this choice as our focus here is to analyze how the CDW and FA transitions are affected by strain near their respective critical temperatures. 

\subsection{Primary CDW mode alone ($L_1^-$)}

We first consider the case of only the primary OP for the CDW state, corresponding to $\vec{\psi}$. The resulting strain-temperature phase diagram, together with its thermodynamic signatures encoded in the ECE and in the specific heat, can be calculated analytically. The free energy, without the secondary order parameter, takes the following simpler form:
\begin{align}
\label{eq:Fpsi}
    \mathcal{F}_\psi=& a_\psi(T-T_\mathrm{CDW})|\vec{\psi}|^2+b_\psi|\vec{\psi}|^4+c_\psi(\psi_1^4+\psi_2^4+\psi_3^4)\\\nonumber
 &+\lambda_{E\psi}\left(\varepsilon_{E1}N_{\psi,1}+\varepsilon_{E2}N_{\psi,2}\right)+\lambda_{A\psi}\varepsilon_A|\vec{\psi}|^2
\end{align}
In the following, we will study how the deviatoric strain $\varepsilon_{E1}$, shear strain $\varepsilon_{E2}$, and the isotropic strain $\varepsilon_A$ affect the $3Q$ CDW transition of the multicomponent order parameter $\vec{\psi}$.

\begin{figure}
    \centering
    \includegraphics[width=\columnwidth]{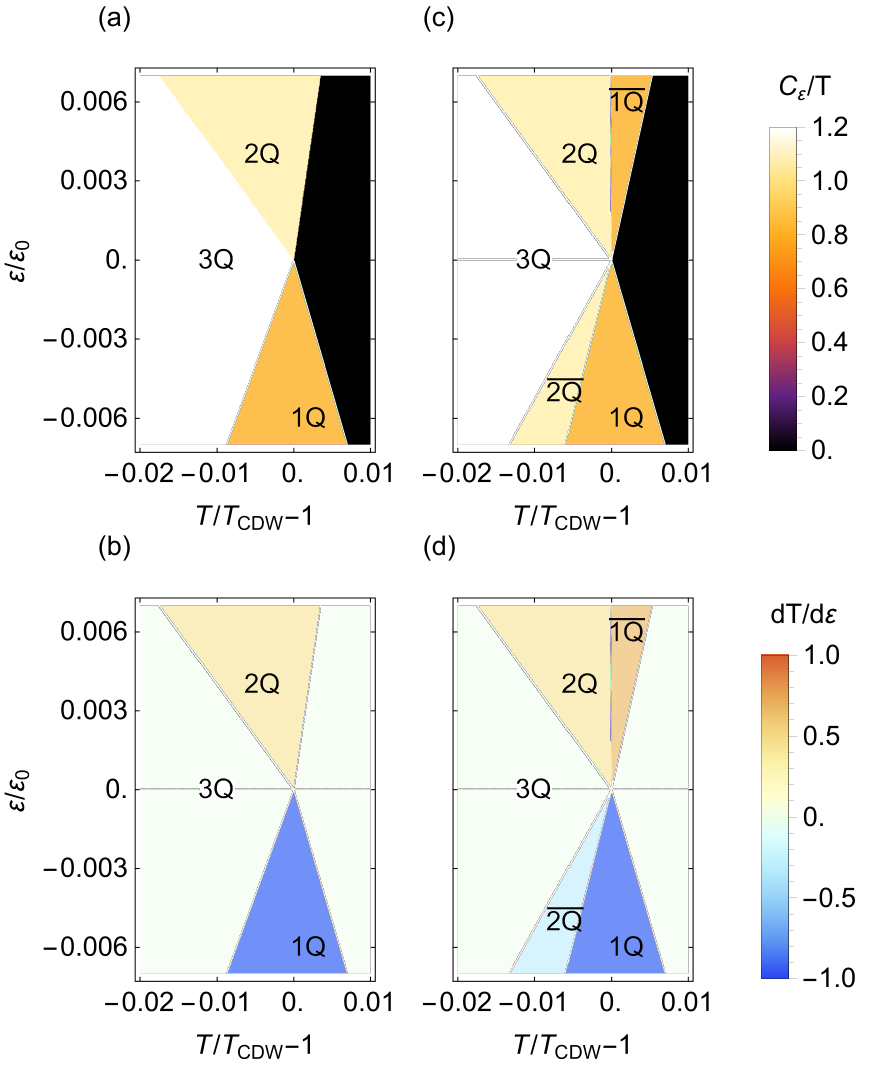}
    \caption{Strain-temperature phase diagram for the primary CDW mode alone [Eq. \eqref{eq:Fpsi}]. (a) Specific heat $C_\varepsilon/T$ [Eqs. \eqref{eq:Ce1Q}-\eqref{eq:Ce3Q}] and (b) ECE coefficient [Eqs. \eqref{eq:ece1Q}-\eqref{eq:ece3Q}] in the case where only the first component of the $E_g$ strain is present, $\varepsilon_E=\varepsilon(1,0)$. The transition temperature lines between different states are given by Eqs. \eqref{eq:T1Q}-\eqref{eq:T2Q3Q}. (c)-(d) Same as (a)-(b) but in the case where the second (shear) component of the $E_g$ strain is also present, $\varepsilon_E=\varepsilon(1,0.3)$. The additional transition lines are given by Eqs.~\eqref{eq:T1Qbar}--\eqref{eq:T1Qbar2Q} and Eqs. \eqref{eq:T1Q2Qbar}--\eqref{eq:T2Qbar3Q}, and the ECE values Eqs.~\eqref{eq:ece1Qbar}--\eqref{eq:ece2Qbar}. In all panels $\varepsilon_0=a_\psi T_\mathrm{CDW}/\lambda_{E\psi}$ and $a_\psi=\lambda_{E\psi}=T_\mathrm{CDW}=4b_\psi=2c_\psi=1$.}
    \label{fig:L1-}
\end{figure}

The coupling to the deviatoric strain $\varepsilon_{E1}$ breaks the threefold degeneracy and drives a bifurcation in the phase diagram. Neglecting the shear component for now, and assuming $\lambda_{E\psi}>0$, compressive strain favors a $1Q$ state $\vec{\psi}_{1Q}=(a,0,0)$, while tensile strain favors a $2Q$ state $\vec{\psi}_{2Q}=(0,a,a)$, as shown in Fig.~\ref{fig:L1-}(a). Such an asymmetry between compressive and tensile strains is characteristic of Potts-nematicity in systems with threefold rotational symmetry \cite{Chakraborty2023}.  Upon further cooling, the system transitions into a modified $3Q$ state with unequal amplitudes $\vec{\psi}_{3Q}=(a,b,b)$ where $a\neq b$. This state is the analog of the threefold symmetric $3Q$ state at zero strain. Since $C_3$ symmetry is not broken spontaneously but by strain, we will not consider $(a,a,a)$ and $(a,b,b)$ as different phases and both will be labeled as $3Q$ in the phase diagram. The order parameter solutions are, respectively, 
\begin{align}
\label{eq:psi1Q}
    |\vec{\psi}_{1Q}|^2&=-\frac{a_\psi(T-T_\mathrm{CDW})+\lambda_{A\psi}\varepsilon_A+\lambda_{E\psi}\varepsilon_{E1}}{2b_\psi+2c_\psi}\\
    |\vec{\psi}_{2Q}|^2&=-\frac{a_\psi(T-T_\mathrm{CDW})+\lambda_{A\psi}\varepsilon_A-\lambda_{E\psi}\varepsilon_{E1}/2}{2b_\psi+c_\psi}\\
    |\vec{\psi}_{3Q}|^2&=-\frac{3a_\psi(T-T_\mathrm{CDW})+3\lambda_{A\psi}\varepsilon_A}{6b_\psi+2c_\psi}.\label{eq:psi3Q}
\end{align}
Note that the $3Q$ order parameter is independent of the $E_g$ component of the strain.
The corresponding critical temperatures for these transitions are, 
\begin{align}
\label{eq:T1Q}
    T_{1Q}&=T_\mathrm{CDW}-\frac{\lambda_{A\psi}}{a_\psi}\varepsilon_A-\frac{\lambda_{E\psi}}{a_\psi}\varepsilon_{E1}\\
    T_{2Q}&=T_\mathrm{CDW}-\frac{\lambda_{A\psi}}{a_\psi}\varepsilon_A+\frac{\lambda_{E\psi}}{2a_\psi}\varepsilon_{E1}\label{eq:T2Q}\\
    T_{1Q-3Q}&=T_\mathrm{CDW}-\frac{\lambda_{A\psi}}{a_\psi}\varepsilon_A+\frac{\lambda_{E\psi}}{2a_\psi}\left(1+\frac{3b_\psi}{c_\psi}\right)\varepsilon_{E1}\\
    T_{2Q-3Q}&=T_\mathrm{CDW}-\frac{\lambda_{A\psi}}{a_\psi}\varepsilon_A-\frac{\lambda_{E\psi}}{a_\psi}\left(1+\frac{3b_\psi}{c_\psi}\right)\varepsilon_{E1},
    \label{eq:T2Q3Q}
\end{align}
and the resulting phase diagram is shown in Fig.~\ref{fig:L1-}(a) for the set of parameters used in Fig. \ref{fig:secIII} for $\mathcal{F}_\psi$. 

The solutions $\vec{\psi}(T,\varepsilon)$ in Eqs.\eqref{eq:psi1Q}-\eqref{eq:psi3Q} allow us to calculate the thermodynamic features across the different phase boundaries. As before, we first neglect the shear component of the strain. Each state has a distinct specific heat value,
\begin{align}
\label{eq:Ce1Q}
    \frac{C_{\varepsilon,1Q}}{T}&=\frac{a_\psi^2}{2b_\psi+2c_\psi}\\
    \frac{C_{\varepsilon,2Q}}{T}&=\frac{a_\psi^2}{2b_\psi+c_\psi}\label{eq:Ce2Q}\\
    \frac{C_{\varepsilon,3Q}}{T}&=\frac{3a_\psi^2}{6b_\psi+2c_\psi}.\label{eq:Ce3Q}
\end{align}
Note that strain selects whether the system is $1Q$, $2Q$ or $3Q$, but the specific heat value is otherwise independent of strain, resulting in plateau features on the phase diagram in Fig.~\ref{fig:L1-}(a), with $C_{\varepsilon,1Q}<C_{\varepsilon,2Q}<C_{\varepsilon,3Q}$ for our choice of parameters $c_\psi=2b_\psi>0$.
Taking the derivative with respect to strain, the ECE coefficient of each state also acquires the following distinct values,
\begin{align}
\label{eq:ece1Q}
    \left(\frac{\partial T}{\partial \varepsilon_{E1}}\right)_{1Q}&=-\frac{\lambda_{E\psi}}{a_\psi}\\
    \left(\frac{\partial T}{\partial \varepsilon_{E1}}\right)_{2Q}&=\frac{\lambda_{E\psi}}{2a_\psi}\label{eq:ece2Q}\\
    \left(\frac{\partial T}{\partial \varepsilon_{E1}}\right)_{3Q}&=0 
    \label{eq:ece3Q}
\end{align}
shown in Fig.~\ref{fig:L1-}(b). As shown in this panel, the ECE is not symmetric with respect to a change in the sign of the strain. Instead, it displays a flat hump feature at tensile strain (orange) and a flat dip with twice the amplitude at compressive strain (blue). 
Note that since the $3Q$ order parameter in Eq. \eqref{eq:psi3Q} is independent of the first $E_g$ component of the strain, the corresponding ECE value in this channel vanishes at this level of approximation. 

An isotropic strain component $\varepsilon_A$, besides uniformly shifting all transition temperatures according to Eqs.\eqref{eq:T1Q}-\eqref{eq:T2Q3Q}, does not affect the specific heat values. Instead, it gives the same additional ECE contribution to all CDW states:
\begin{equation}
    \frac{\partial T}{\partial \varepsilon_A}=-\frac{\lambda_{A\psi}}{a_\psi}.
\end{equation}

We now study the effect of a finite shear component $\varepsilon_{E2}\neq 0$, which may arise from sample misalignment when strain is intended along a principal axis, or from uniaxial stress deliberately applied along a direction rotated with respect to the crystallographic axis \cite{Ian2026}. 
A finite shear component explicitly breaks the remaining $C_{2x}$ and mirror symmetries and modifies the phase diagram by affecting two of the transitions, as shown in Fig. \ref{fig:L1-}(c). In the following, we take $\varepsilon_{E2}$ to have the same sign as $\varepsilon_{E1}$; the opposite case is equivalent by symmetry, with the roles of the second and third components of $\vec\psi$ exchanged. At tensile strain, the single $2Q$ CDW transition becomes a double $\overline{1Q}$--$2Q$ CDW transition with the following order parameter for the new $\overline{1Q}$ state, 
\begin{equation}
    |\vec{\psi}_{\overline{1Q}}|^2=|\vec{\psi}_{1Q}|^2+\lambda_{E\psi}\frac{\frac{3}{2}\varepsilon_{E1}+\frac{\sqrt{3}}{2}\varepsilon_{E2}}{2b_\psi+2c_\psi}.
\end{equation}
Meanwhile, $|\vec{\psi}_{2Q}|^2$ remains unchanged, and the critical temperatures are modified to
\begin{align}
    \label{eq:T1Qbar}T_{\overline{1Q}}&=T_{2Q}+\frac{\lambda_{E\psi}}{a_\psi}\frac{\sqrt{3}}{2}\varepsilon_{E2}\\
    T_{\overline{1Q}-2Q}&=T_{2Q}-\frac{\lambda_{E\psi}}{a_\psi}\frac{\sqrt{3}}{2}\left(1+\frac{2b_\psi}{c_\psi}\right)\varepsilon_{E2}.\label{eq:T1Qbar2Q}
\end{align}
Similarly, at compressive strain the $1Q$--$3Q$ transition splits into $1Q$--$\overline{2Q}$--$3Q$ transitions, with 
\begin{equation}    |\vec{\psi}_{\overline{2Q}}|^2=|\vec{\psi}_{2Q}|^2-\lambda_{E\psi}\frac{\frac{3}{2}\varepsilon_{E1}+\frac{\sqrt{3}}{2}\varepsilon_{E2}}{2(2b_\psi+c_\psi)}
\end{equation}
The $3Q$ CDW OP $|\vec{\psi}_{3Q}|^2$ remains unchanged, and the critical temperatures become:
\begin{align}
\label{eq:T1Q2Qbar}
    T_{1Q-\overline{2Q}}&=T_{1Q-3Q}-\frac{\lambda_{E\psi}}{a_\psi}\frac{\sqrt{3}}{2}\left(1+\frac{b_\psi}{c_\psi}\right)\varepsilon_{E2}\\
    T_{\overline{2Q}-3Q}&=T_{1Q-3Q}+\frac{\lambda_{E\psi}}{a_\psi}\frac{\sqrt{3}}{2}\left(1+\frac{3b_\psi}{c_\psi}\right)\varepsilon_{E2}.\label{eq:T2Qbar3Q}
\end{align}

Since the shear strain only shifts $|\vec\psi|^2$ by a temperature independent amount, the specific heat still depends only on whether the state is $1Q$, $2Q$ or $3Q$ [Eqs. \eqref{eq:Ce1Q}-\eqref{eq:Ce3Q}]. In particular, the new $\overline{1Q}$ phase at tensile strain and the $\overline{2Q}$ phase at compressive strain have specific heat expressions identical to those of the $1Q$ and $2Q$ states, respectively [Fig. \ref{fig:L1-}(c)].

The ECE, by contrast, acquires new values in these two states. For strain along $\varepsilon_E=\varepsilon(1,r)$:
\begin{align}
\label{eq:ece1Qbar}
     \left(\frac{\partial T}{\partial \varepsilon}\right)_{\overline{1Q}}&=\frac{\lambda_{E\psi}}{2a_\psi}\left(1+\sqrt{3}r\right)\\
     \left(\frac{\partial T}{\partial \varepsilon}\right)_{\overline{2Q}}&=-\frac{\lambda_{E\psi}}{4a_\psi}\left(1+\sqrt{3}r\right)\label{eq:ece2Qbar}
\end{align}
The remaining states retain the same values as in the case without shear strain, given by Eqs. \eqref{eq:ece1Q}-\eqref{eq:ece3Q}, and shown in Fig. \ref{fig:L1-}(d).

\subsection{Coexistence of primary and secondary CDW modes ($L_1^-+L_2^-$)}

We study now how the phase diagram and thermodynamic features are modified when the system undergoes an additional FA transition through the condensation of a secondary CDW OP, $\vec{\eta}$, at a lower temperature $T_\mathrm{FA}<T_\mathrm{CDW}$ [Fig. \ref{fig:secIII}(a)]. Including the secondary order parameter $\vec{\eta}$ generates several additional terms in the free energy, as discussed in Section \ref{sec:GL} and Eq. \eqref{eq:Feps}, which makes obtaining analytical expressions more challenging. To simplify the discussion, we consider only the case where there is no shear strain component present,  $\varepsilon_{E}=\varepsilon(1,0)$, to assess the effect of FA order on the phase diagram and thermodynamic features shown in Figs. \ref{fig:L1-}(a)-(b). We note however that shear strain effects will be analogous to those discussed for the primary mode, and can be in principle computed in a straightforward way. 

\begin{figure}[h!]
    \centering    \includegraphics[width=\columnwidth]{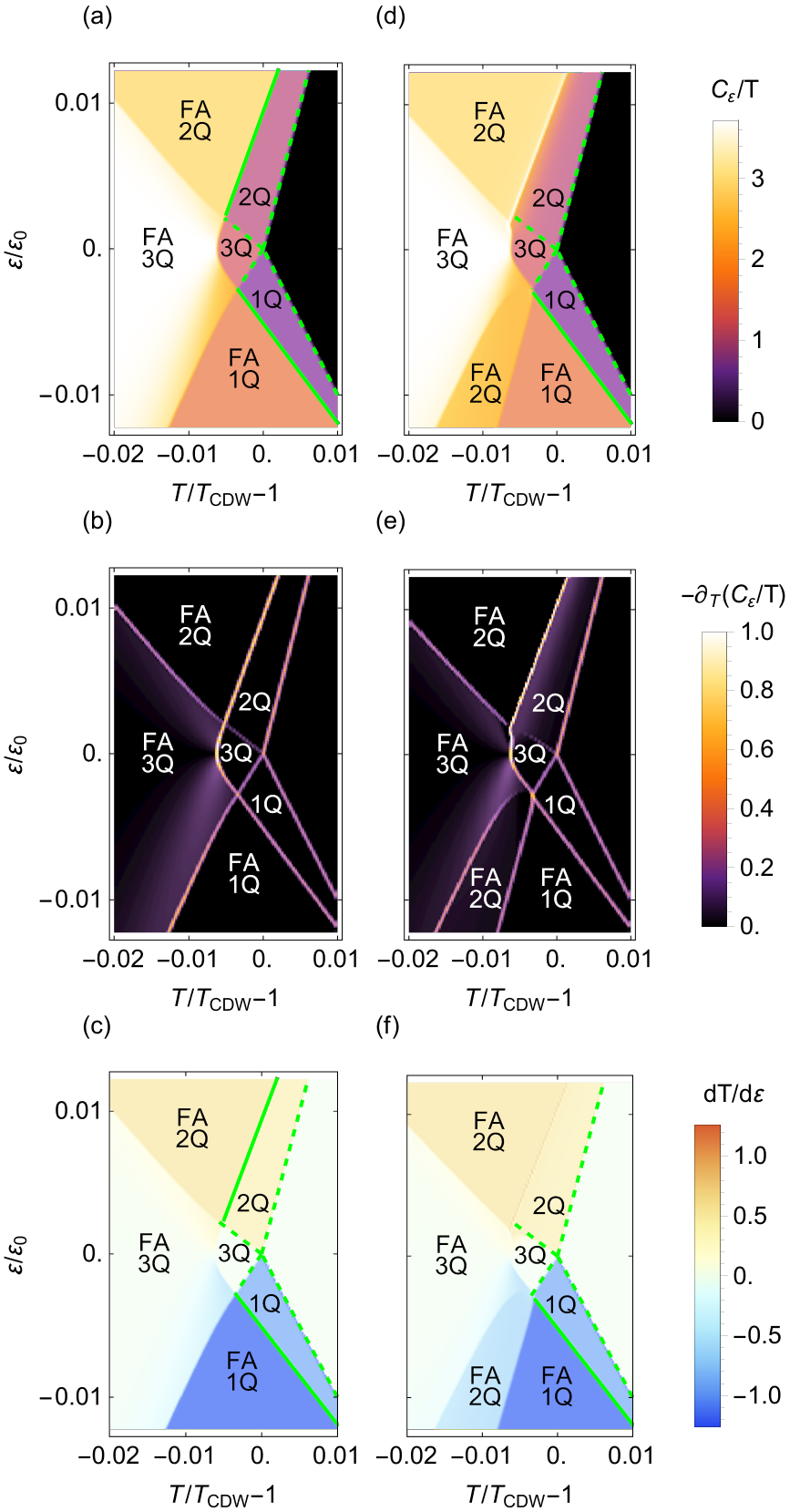}
    \caption{Strain-temperature phase diagram for the coexistence of the primary and secondary CDW modes in the presence of the first component of the $E_g$ strain, $\varepsilon_E=\varepsilon(1,0)$. Numerical solutions of (a) the specific heat $C_\varepsilon/T$, (b) its temperature derivative $-\partial_T(C_\varepsilon/T)$, and (c) the ECE for $d_1=-0.4$. (d)-(f) Same as (a)-(c) for $d_1=-d_4=-0.4$. A qualitatively similar phase diagram is obtained for $\lambda_{E\psi\eta}>0$ (both with $d_4=0$ and $d_4>0$), consistent with experimental data in \cite{sarah2026}.
   The analytical transition lines are highlighted in green, with the dashed lines given by Eqs.\eqref{eq:T1Q}-\eqref{eq:T2Q3Q} and the full lines by Eq. \eqref{eq:T1QFA} and Eq.\eqref{eq:T2QFA}. In all panels, $a_\psi=T_\mathrm{CDW}=\lambda_{E\psi}=1$, $T_\eta/T_\mathrm{CDW}=0.99$, $\lambda_{E\psi\eta}=0$, $a_\eta=0.8a_\psi$, $b_\psi=b_\eta=0.25$, $c_\psi=c_\eta=0.5$ and $\lambda_{E\eta}=1.3\lambda_{E\psi}$.}
    \label{fig:L1-L2-}
\end{figure}

\begin{figure*}[]
    \centering
\includegraphics[width=0.85\linewidth]{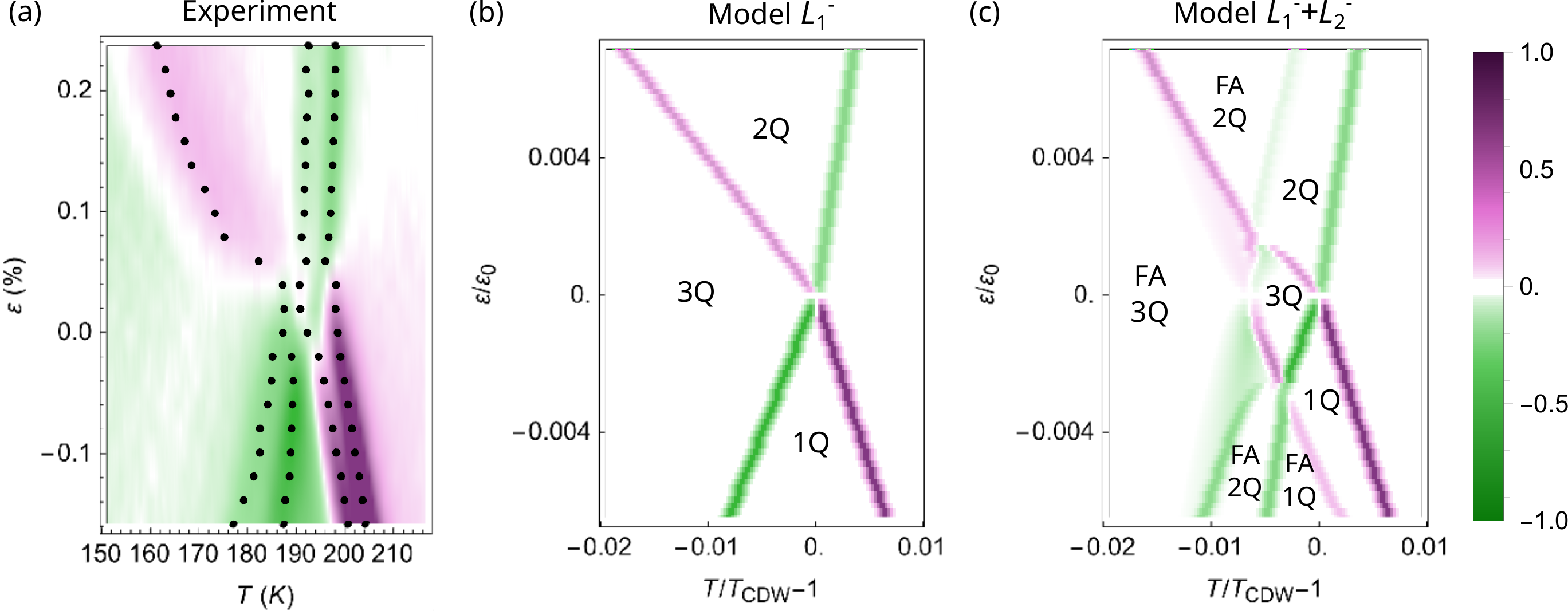}
    \caption{Comparison between theoretical and experimental results for the temperature derivative of the ECE, $\partial_T (\partial T/\partial \varepsilon)$ . (a) Experimental data adapted from \cite{sarah2026}. The black dots indicate the peaks and dips. Results from the Landau model with (b) primary mode $L_1^-$ only (derivative of ECE in Fig. \ref{fig:L1-}(b)) and (c) primary and secondary modes $L_1^-+L_2^-$ (derivative of ECE in Fig. \ref{fig:L1-L2-}(f)). For visualization, the model transitions are broadened by a Gaussian of width $\sigma=4 \times 10^{-4}$ in $T/T_\mathrm{CDW}$. Each data set is normalized to its most negative value, and all panels share the same colour scale, clipped at $\pm1$.}
    \label{fig:exp}
\end{figure*}

We start on the compressive strain side ($\varepsilon<0$), where a $1Q$ solution is favored with $\psi_1\neq0$ at $T_{1Q}$ [Eqs. \eqref{eq:psi1Q} and \eqref{eq:T1Q}]. At temperatures below $T_{1Q}$, the secondary mode condenses, $\eta_1\neq 0$, and a FA transition takes place within the $1Q$ phase, as shown in Fig. \ref{fig:L1-L2-}(a). The equilibrium solutions for $\psi_1$ and $\eta_1$ 
under strain depend only on the parameters $d_1$, $d_3$ and $\lambda_{E\nu}$,
\begin{align}
    |\vec{\psi}_{1Q,\mathrm{FA}}|^2&=\frac{\vec{\psi}_{1Q}^2-\frac{a_\eta}{a_\psi}D_{1\psi}\vec{\eta}_{1Q}^2}{1-D_{1\psi}D_{1\eta}}\\
    |\vec{\eta}_{1Q,\mathrm{FA}}|^2&=\frac{\vec{\eta}_{1Q}^2-\frac{a_\psi}{a_\eta}D_{1\eta}\vec{\psi}_{1Q}^2}{1-D_{1\psi}D_{1\eta}}\\
D_{1\nu}&=\frac{a_\nu(d_1+d_3)}{a_{\bar{\nu}}(2b_\nu+2c_\nu)}\label{eq:D1nu}
\end{align} 
where we have defined the analogue of Eq. \eqref{eq:psi1Q} for the secondary mode when uncoupled from the primary mode, $|\vec{\eta}_{1Q}|^2=-\frac{a_\eta(T-T_\eta)+\lambda_{E\eta}\varepsilon_{E1}}{2b_\eta+2c_\eta}$. Here $\bar{\nu}$ denotes the complementary order parameter, i.e., $\bar{\psi}=\eta$ and $\bar{\eta}=\psi$. From the condition $|\vec{\eta}_{1Q,\mathrm{FA}}|^2=0$ we obtain the $1Q$ CDW to $1Q$ FA transition temperature, 
\begin{equation}
\label{eq:T1QFA}
T_{1Q,\mathrm{FA}}=\frac{T_\eta-D_{1\psi}T_\mathrm{CDW}-\left(\frac{\lambda_{E\eta}}{a_\eta}-D_{1\psi}\frac{\lambda_{E\psi}}{a_\psi}\right)\varepsilon_{E1}}{1-D_{1\psi}} 
\end{equation}
which shows a linear behavior as a function of strain, as seen in Fig. \ref{fig:L1-L2-}(a). 
Experimentally, the CDW transition $T_{1Q}$ [Eq. \eqref{eq:T1Q}] and the FA transition $T_{1Q,\mathrm{FA}}$ are approximately parallel to each other as a function of strain  \cite{sarah2026} (see also Fig. \ref{fig:exp}(a)). To understand the conditions under which our model gives this behavior, we consider the regime of small coupling parameters $d_i$, $|D_{1\nu}|\ll 1$. Then, Eq.\eqref{eq:T1QFA} can be expanded to leading order in $D_{1\nu}$ yielding
\begin{equation}
\label{eq:T1QFAlim}
     T_{1Q,\mathrm{FA}}\approx T_\eta-\frac{\lambda_{E\eta}}{a_\eta}\varepsilon_E
     +\mathcal{O}(D_{1\nu})
\end{equation}
Thus, requiring both transitions to be approximately parallel as a function of strain imposes the following constraints on the model parameters: $\mathrm{sgn}(\lambda_{E\eta})=\mathrm{sgn}(\lambda_{E\psi})$ and $\frac{\lambda_{E\eta}}{a_\eta}\approx \frac{\lambda_{E\psi}}{a_\psi}$. The exact condition will be discussed later.

We turn now to the tensile strain side of the phase diagram ($\varepsilon>0$).
Since the parameters $\lambda_{E\psi\eta}$, $d_4$ and $d_5$ couple to odd powers of the components of $\vec\eta$ in the $2Q$ state, the mean-field equations no longer reduce to linear equations for the squared amplitudes. We therefore set these parameters to zero to obtain analytical expressions, and solve the general case numerically. At temperatures below $T_{2Q}$, the secondary mode condenses with $\eta_2=\eta_3\neq 0$ and a FA transition takes place,
\begin{align}
\vec{\psi}_{2Q,\mathrm{FA}}^2&=\frac{\vec{\psi}_{2Q}^2-\frac{a_\eta}{a_\psi}D_{2\psi}\vec{\eta}_{2Q}^2}{1-D_{2\psi}D_{2\eta}} \\
   \vec{\eta}_{2Q,\mathrm{FA}}^2&= \frac{\vec{\eta}_{2Q}^2-\frac{a_\psi}{a_\eta}D_{2\eta}\vec{\psi}_{2Q}^2}{1-D_{2\psi}D_{2\eta}}\\
D_{2\nu}&=\frac{a_\nu(d_1+d_3+d_2/4)}{a_{\bar\nu}(2b_\nu+c_\nu)}\label{eq:D2nu}
\end{align} 
Note that, unlike the case of the $1Q$-FA solution, the order parameters depend explicitly on the parameter $d_2$. From the condition $\vec{\eta}_{2Q,\mathrm{FA}}^2=0$ we obtain the $2Q$-FA transition temperature,
\begin{equation}
\label{eq:T2QFA}
T_{2Q,\mathrm{FA}}=\frac{T_\eta-D_{2\psi}T_\mathrm{CDW}+\left(\frac{\lambda_{E\eta}}{2a_\eta}-D_{2\psi}\frac{\lambda_{E\psi}}{2a_\psi}\right)\varepsilon_{E1}}{1-D_{2\psi}} 
\end{equation}
Similarly to the compressive strain side of the phase diagram, on the tensile side the critical temperature $T_{2Q,\mathrm{FA}}$ also evolves linearly with strain (green full line in Fig. \ref{fig:L1-L2-}(a)). However, it has a different slope compared to $T_{1Q,\mathrm{FA}}$, mirroring the compressive-tensile asymmetry found for the primary mode.
In fact, in the limit of small interacting parameters, $|D_{2\nu}|\ll 1$, the FA transition reduces to leading order
\begin{equation}
\label{eq:T2QFAlim}
    T_{2Q,\mathrm{FA}}\approx T_\eta+\frac{\lambda_{E\eta}}{2a_\eta}\varepsilon_{E}+\mathcal{O}(D_{2\nu}),
\end{equation}
and we recover the same asymmetry with a factor of $2$ difference between $T_{1Q,\mathrm{FA}}$ [Eq. \eqref{eq:T1QFAlim}] and $T_{2Q,\mathrm{FA}}$ [Eq. \eqref{eq:T2QFAlim}], see Eqs. \eqref{eq:T1Q}-\eqref{eq:T2Q}. 

We can now obtain the thermodynamic features at these two ferroaxial transitions encoded in the specific heat and in the ECE.
Defining the weights 
\begin{align}
    w_{1Q,\nu}&\equiv\frac{C_{\varepsilon,1Q,\nu}}{T}\left(1-D_{1\bar\nu}\right)\label{eq:w1Q}\\
    w_{2Q,\nu}&\equiv\frac{C_{\varepsilon,2Q,\nu}}{T}\left(1-D_{2\bar\nu}\right)\label{eq:w2Q}
\end{align}
for $\nu=\psi,\eta$, with $C_{\varepsilon,1Q,\nu}$ and $C_{\varepsilon,2Q,\nu}$ given by Eqs.~\eqref{eq:Ce1Q}--\eqref{eq:Ce2Q}, the specific heat at constant strain in the FA phases reads
\begin{align}
  \label{eq:Ce1QFA}
    \frac{C_{\varepsilon,1Q,\mathrm{FA}}}{T}&=\frac{w_{1Q,\psi}+w_{1Q,\eta}}{1-D_{1\psi} D_{1\eta}}\\
    \frac{C_{\varepsilon,2Q,\mathrm{FA}}}{T}&=\frac{w_{2Q,\psi}+w_{2Q,\eta}}{1-D_{2\psi} D_{2\eta}}\label{eq:Ce2QFA}
\end{align}
The specific heat in the FA phase is thus built from the contributions of the two order parameters, each weighted by the factor $1-D_{\bar\nu}$ that accounts for its coupling to the other, and collectively renormalized by $1/(1-D_{i\psi} D_{i\eta})$ with $i=1,2$. 
Relative to the uncoupled primary mode, the specific heat jumps at the FA transition by,
\begin{align}
\frac{C_{\varepsilon,1Q,\mathrm{FA}}-C_{\varepsilon,1Q}}{T}&=\frac{C_{\varepsilon,1Q,\eta}}{T}\frac{(1-D_{1\psi})^2}{1-D_{1\psi}D_{1\eta}}\label{eq:DC1Q}\\
\frac{C_{\varepsilon,2Q,\mathrm{FA}}-C_{\varepsilon,2Q}}{T}&=\frac{C_{\varepsilon,2Q,\eta}}{T}\frac{(1-D_{2\psi})^2}{1-D_{2\psi}D_{2\eta}}\label{eq:DC2Q}
\end{align}
A stable FA transition requires $D_{i\psi}<1$, since otherwise $\vec \eta$ does not condense, and $D_{i\psi} D_{i\eta}<1$ for the free energy to be bounded. Within this range the jump is always positive. 
In the absence of coupling ($D_{1\nu}=D_{2\nu}=0$), it reduces to the specific heat of the uncoupled secondary mode, whereas an attractive coupling $D_{\nu}<0$ enhances it. 

As illustrated in Fig. \ref{fig:L1-L2-}(a) by a numerical calculation for $d_1=-0.4$ and the other $d_i=0$ (corresponding to Fig.~\ref{fig:secIII}(a) for the zero-strain case), the specific heat jumps at the FA transition $T_{1Q,\mathrm{FA}}$ and $T_{2Q,\mathrm{FA}}$ are significant and comparable in magnitude. At even lower temperatures, the system transitions into a $3Q$ state, which is now also ferroaxial, through another jump in the specific heat, similarly to the phase diagram of the primary mode [Fig. \ref{fig:L1-}(a)]. The temperature derivative of the specific heat, shown in Fig. \ref{fig:L1-L2-}(b), clearly reflects the hierarchy of transitions taking place in the presence of strain, with three transition temperatures for finite strain. Experimentally, however, an additional transition is found on the compressive side [Fig.~\ref{fig:exp}(a)]; we will return to this point at the end of this section.   

The behavior of the ECE can also be obtained analytically at the FA transition on the compressive and tensile sides. 
The ECE at the FA transitions is a weighted mean of the contributions of the two order parameters, with the weights of Eqs.~\eqref{eq:w1Q}--\eqref{eq:w2Q},
\begin{align}
    \left(\frac{\partial T}{\partial \varepsilon_{E1}}\right)_{1Q,\mathrm{FA}}&=\frac{\sum_{\nu}w_{1Q,\nu}\left(\frac{\partial T}{\partial \varepsilon_{E1}}\right)_{1Q,\nu}}{\sum_\nu w_{1Q,\nu}}\label{eq:ece1QFA}\\
    \left(\frac{\partial T}{\partial \varepsilon_{E1}}\right)_{2Q,\mathrm{FA}}&=\frac{\sum_{\nu}w_{2Q,\nu}\left(\frac{\partial T}{\partial \varepsilon_{E1}}\right)_{2Q,\nu}}{\sum_\nu w_{2Q,\nu}}\label{eq:ece2QFA}
\end{align}
where the ECE of the uncoupled order parameters is given by Eqs.~\eqref{eq:ece1Q}--\eqref{eq:ece2Q}.
The jump of the ECE at the FA transitions follows directly,
\begin{align}
\Delta_{1Q,\mathrm{FA}}&=-\frac{w_{1Q,\eta}}{w_{1Q,\psi}+w_{1Q,\eta}}\left(\frac{\lambda_{E\eta}}{a_\eta}-\frac{\lambda_{E\psi}}{a_\psi}\right),\label{eq:jump1Q}\\
\Delta_{2Q,\mathrm{FA}}&=\frac{w_{2Q,\eta}}{2(w_{2Q,\psi}+w_{2Q,\eta})}\left(\frac{\lambda_{E\eta}}{a_\eta}-\frac{\lambda_{E\psi}}{a_\psi}\right).\label{eq:jump2Q}
\end{align}
Since the weight fractions are positive whenever the FA phase exists, the jumps on the compressive and tensile sides always have opposite signs, both controlled by the sign of $\lambda_{E\eta}/a_\eta-\lambda_{E\psi}/a_\psi$.
Figure \ref{fig:L1-L2-}(c) shows the corresponding ECE signal for the case shown in Fig.~\ref{fig:L1-L2-}(a)-(b) ($d_1=-0.4$), where $\lambda_{E\eta}/a_\eta>\lambda_{E\psi}/a_\psi$ and the jumps are negative on the compressive side and positive on the tensile side, in agreement with the experimental results \cite{sarah2026}.

Experimentally $T_{1Q,\mathrm{FA}}(\varepsilon)$ is also approximately parallel to $T_{1Q}(\varepsilon)$ \cite{sarah2026}. From Eq. \eqref{eq:T1Q} and Eq. \eqref{eq:T1QFA} their slopes differ by
\begin{equation}
    \frac{dT_{1Q,\mathrm{FA}}}{d\varepsilon_{E1}}-\frac{dT_{1Q}}{d\varepsilon_{E1}}=-\frac{1}{1-D_{1\psi}}\left(\frac{\lambda_{E\eta}}{a_\eta}-\frac{\lambda_{E\psi}}{a_\psi}\right),
\end{equation}
i.e. by the same combination that controls the ECE jumps [Eqs.~\eqref{eq:jump1Q}--\eqref{eq:jump2Q}]. Exactly parallel lines would therefore imply no jump. Both features can be nevertheless reconciled qualitatively for $\lambda_{E\eta}/a_\eta$ moderately larger than $\lambda_{E\psi}/a_\psi$, since an attractive coupling reduces the corresponding slope mismatch by the factor $1/(1-D_{1\psi})$. This motivates the choice $a_\eta=0.8a_\psi$ and $\lambda_{E\eta}=1.3\lambda_{E\psi}$ in Fig. \ref{fig:L1-L2-}.

Finally, we show the strain-temperature phase diagram and its thermodynamic features in Fig. \ref{fig:L1-L2-}(d)-(f) for  $d_1=-d_4=-0.4$. These values still realize Case 1 at zero strain [Fig. \ref{fig:secIII}(a)], since the $d_4$ term vanishes in the $C_3$-symmetric $3Q$ state and is too weak to destabilize it, but becomes active once strain makes the order-parameter components unequal. On the tensile side, $d_4$ induces a finite $\eta_3-\eta_2$ already in the $2Q$ state, so that the subsequent FA-$2Q$ transition has no closed-form expression and, for sufficiently large $|d_4|$, becomes first order, as in Fig. \ref{fig:L1-L2-}(d)-(f). On the compressive side, the phase diagram now exhibits an additional transition in all three panels, as the ferroaxial $1Q-3Q$ transition splits, with an intermediate FA $2Q$ state appearing in between. This 2Q state corresponds to a $(\nu_1,\nu_2,0)$ configuration, with $\nu_1 \neq \nu_2$ for both OPs ($\nu=\psi,\eta$). Thus, it can appear only once the $C_{2x}$ symmetry relating the second and third components of the OPs is broken by FA order, through the parameters $d_4$, $d_5$ [Eq. \eqref{eq:FL1L2}] or the strain coupling $\lambda_{E\psi\eta}$ [Eq. \eqref{eq:Feps}],  which generate terms antisymmetric in $\nu_2$ and $\nu_3$. Indeed, qualitatively similar results are also found for $d_1=-0.4$ and $\lambda_{E\psi\eta}>0$. Interestingly, the ECE measurements of Ref. \cite{sarah2026} reveal not three but four transition lines on the compressive side of the strain phase diagram, consistent with the existence of this FA-$2Q$ state. 

We finish the section by comparing in Fig.~\ref{fig:exp} the temperature derivative of the elastocaloric coefficient, $\partial_T(dT/d\varepsilon)$, measured experimentally in Ref.~\cite{sarah2026} [panel~(a)], with our theoretical results for a single primary order parameter $L_1^-$ [panel~(b)], and for the coexistence of primary and secondary modes $L_1^- + L_2^-$ [panel~(c)]. As shown in this figure, the single-mode calculation reproduces only a subset of the features seen experimentally, whereas the additional condensation of $L_2^-$ captures all the observed features remarkably well.

\section{Conclusions}
\label{sec:conclusions}
In this work, we shed light on the sequence of phase transitions experimentally observed in TiSe$_2$. While the high-temperature CDW transition at $T_\mathrm{CDW}\approx 200$ K is well understood as arising from the condensation of the $L_1^-$ mode \cite{disalvo1976,holt2001,ueda2021}, the subsequent ferroaxial and nematic transitions, recently observed at $T_\mathrm{FA}\approx 193$ K and $T_\mathrm{nem}\approx 165$ K \cite{sarah2026,Ian2026,Lv26}, call for a secondary mode that condenses at lower temperature and couples to $L_1^-$. We set up a Landau model of two coupled three-component order parameters, with $L_1^-$ as the primary one. Our results show that the experimentally observed hierarchy of transitions $T_\mathrm{CDW} > T_\mathrm{FA} > T_\mathrm{nem}$ is naturally realized when the secondary mode is $L_2^-$. This mode is one among several soft modes seen in DFT whose frequencies are thermally accessible at the CDW transition temperature~\cite{Bianco15,Hellgren17,KimUnpublished} and its condensation energy was reported to be comparable to $L_1^-$ \cite{Wegner2020}. Indeed, this mode allows for a gap opening between the folded conduction and valence $p_x,p_y$-derived bands, like the primary $L_1^-$ mode, as well as with a $p_z$-derived valence band  that the primary mode affects only via spin-orbit coupling. Recent diffraction experiments also revealed finite but weak $(h0l)$ peaks~\cite{kim2024}, which implies the glide extinction of the $L_1^-$ mode is weakly broken, compatible with an additional longitudinal mode like $L_2^-$. 

The experimental strain-temperature phase diagram of TiSe$_2$ provides additional evidence for the presence of a secondary order parameter~\cite{sarah2026}. The temperature derivative of the elastocaloric coefficient shown in Fig.~\ref{fig:exp}(a) cannot be fully reproduced by the condensation of the $L_1^-$ mode alone [Fig.~\ref{fig:exp}(b)]. In this work, we showed that, by including the condensation of $L_2^-$ as a secondary mode, all the experimentally observed features of $\partial_T (dT/d\varepsilon)$ can be reproduced [Fig.~\ref{fig:exp}(c)]. Our theoretical model thus accounts for all the elastocaloric coefficient anomalies, in terms of the sequences of phase transitions $1Q \to \text{FA-}1Q \to \text{FA-}2Q \to \text{FA-}3Q$ on the compressive side, and $2Q \to \text{FA-}2Q \to \text{FA-}3Q$ on the tensile side.

Interestingly, a coexistence of order parameters with opposite $\sigma_d$ parity has also been recently observed in monolayer samples in Ref. \cite{Wan26}. That work assumed that, in addition to the standard CDW order parameter proposed by Di Salvo \emph{et al.} \cite{disalvo1976}, which becomes $M_1^-$ in the monolayer case, there was an additional $\sigma_d$-even order parameter with symmetry $M_1^+$. However, the monolayer version of the $L_2^-$ order parameter proposed in our work is  $M_2^-$, which is indistinguishable from $M_1^+$ in STM which mostly probes the top layer and is therefore not sensitive to inversion. An $M_2^-$ order parameter could also represent an alternative explanation of these experiments.

This scenario, in which the ferroaxial state preserves inversion symmetry, may at first sight appear to be at odds with the observed circular photogalvanic effect (CPGE), a vertical photocurrent that switches sign when the circular polarization of light is reversed ~\cite{Xu20}. This photocurrent is only found when the CDW is prepared by cooling in the presence of circularly polarized light (i.e., with training before the photocurrent measurement). While normally CPGE is taken as a signature of inversion symmetry breaking, in this particular experimental setup it can actually be explained both by chiral and FA order as follows. In the chiral scenario, the training field for chirality is $q_z\vec E\times\vec E^*$, and the photocurrent is $J_z \propto (\vec E\times\vec E^*)_z$. In the FA scenario, the training field is $\vec E\times\vec E^*$ and the photocurrent can be explained through the photon drag effect $J_z \propto q_z(\vec E\times\vec E^*)_z$. Both explanations require a finite $q_z$ of light and could in principle be differentiated by comparing the training and measurement with forward and backward illumination, where $q_z$ is reversed. In addition, these ideal scenarios are complicated by the facts that vertical photocurrent measurements can occur even in centrosymmetric systems, because the top contact is irradiated more strongly than the bottom one, and that it is not feasible to irradiate at the midpoint to ensure photothermal effects can be discarded. A contactless experiment that probes the THz photocurrent emitted from pulsed light might also help discern between the two scenarios. Importantly, recent second-harmonic generation data found no evidence of inversion symmetry breaking in TiSe$_2$ \cite{Ian2026}, consistent with the ferroaxial scenario and at odds with the chiral scenario. 

More broadly, the framework developed here where multi-component CDW phase transitions are probed with thermodynamic measurements in the presence of strain is not specific to TiSe$_2$ and can be readily applied to other CDW materials hosting multi-component order parameters. 
Promising candidates include kagome metals of the $A$V$_3$Sb$_5$ family, where the interplay between $1Q$, $2Q$, and $3Q$ charge order, nematicity, and possible chiral or ferroaxial signatures has been the subject of intense debate \cite{Christensen2021,Ritz2023,Sante2026}, as well as other trigonal or hexagonal transition-metal dichalcogenides exhibiting multi-$Q$ CDW order, such as 1T-TaS$_2$ and 2H-NbSe$_2$~\cite{Wilson1975}. Our results suggest that, in such systems, the symmetry breaking patterns due to coupled multi-Q modes can be effectively uncovered by symmetry-guided Landau theories of the strain induced thermodynamic and transport responses. 

Finally, we note that our analysis is restricted to a homogeneous Landau theory and therefore neglects the role of spatial fluctuations. A study of the full Ginzburg--Landau model, including gradient terms, would help clarify the nature of the different phase transitions discussed here, and could further reveal the possible emergence of new intermediate phases stabilized by fluctuations.

\section*{Acknowledgements}

We thank Sarah Edwards and Jiun-Haw Chu for sharing with us the ECE data from Ref.~ \cite{sarah2026} shown in Figure \ref{fig:exp}. We thank Liwia Karakula, T. Birol, E. Day-Roberts and I. R. Fisher for fruitful discussions. M.N.G. is supported by the grants RYC2021-031639-I and PID2023-153277NB-I00 funded by MCIN/AEI/10.13039/501100011033. F. J. is supported by grant PID2025-175470NB-I00 from the same institution and by a 2024 Leonardo Grant for Scientific Research and Cultural Creation, BBVA Foundation. I.M. acknowledges financial support by the Swiss National Science Foundation (SNSF) via the SNSF postdoctoral Grant No. TMPFP2\_217204.

\appendix

\section{Landscape of states from $L_1^-+L_2^-$}
\label{app:di}

Table~\ref{tab:di} lists the possible states that each coupling coefficient $d_i$ favors in isolation. The results are obtained by numerical minimization of the free energy with a single nonzero $d_i$ in Eq. \eqref{eq:FL1L2}. 

\begin{table*}[]
    \centering
    \begin{tabular}{c|c|c|c|c|l}
         & $(\psi_1,\psi_2,\psi_3)$ & $(\eta_1,\eta_2,\eta_3)$ &  FA $\parallel c$ & Nematic, FA $\perp c$ & Space group \\\hline
        $d_1>0$ & 
         $(a,b,b)$ & $(-a',b',b')$ & yes & yes & {\color{magenta}$P\bar 1$}\\\hline
         $d_1<0$ & $(a,a,a)$ & $(a',a',a')$ & yes & no & {\color{RoyalPurple}$P\bar 3$}\\\hline
         $d_2>d_2^*>0$ & $(a,a,b)$ $\to$ $(a,a,0)$ & $(-a',a',0)$ & no  & yes  & {\color{PineGreen}$C2/c$} $\to$ {\color{LimeGreen}$C2/m$}\\\hline
         $d_2^*>d_2>0$ & $(a,b,b)$ & $(-a',b',b')$ & yes  & yes & {\color{magenta}$P\bar 1$} \\\hline
         $d_2<0$ & $(a,a,a)$ & $(a',a',a')$ & yes & no & {\color{RoyalPurple}$P\bar 3$} \\\hline         
         $d_3\neq 0$ & $(a,a,a)$ & $(a',a',\pm a')$ & yes  & no/yes & {\color{RoyalPurple}$P\bar 3$}/{\color{magenta}$P\bar 1$}\\\hline
         $|d_4|>d_4^*$ & $(a,a,b)$ $\to$ $(a,a,0)$ & $(-a',a',0)$ $\to$ $(-a',a',b')$ & no & yes  & {\color{PineGreen}$C2/c$} $\to$ {\color{LimeGreen}$C2/m$} \\\hline
          $|d_4|<d_4^*$ & $(a,a,b)$ $\to$ $(a,b,c)$ & $(a',-a',0)$ $\to$ $(b',-a',c')$ & no $\to$ yes & yes & {\color{PineGreen}$C2/c$} $\to$ {\color{magenta}$P\bar 1$}$^\dagger$ \\\hline
         $|d_5|>d_5^*$ & $(a,a,b)$ & $(-a',a',0)$ & no  & yes  & {\color{PineGreen}$C2/c$}\\\hline
         $|d_5|<d_5^*$ & $(a,a,b)$ $\to$ $(a,b,c)$ & $(-a',a',0)$ $\to$ $(-b', a', c')$   & no $\to$ yes & yes & {\color{PineGreen}$C2/c$} $\to$ {\color{magenta}$P\bar 1$}\\\hline
    \end{tabular}
    \caption{The coexistence state favored below the 3Q CDW transition with $ (\psi_1,\psi_2,\psi_3) $ and $(\eta_1,\eta_2,\eta_3)$, obtained by considering each coupling parameter $d_i$ in Eq.~\eqref{eq:FL1L2} in an isolated way. The remaining parameters entering Eqs.~\eqref{eq:FL1}-\eqref{eq:FL2} are $a_\psi=a_\eta=T_\mathrm{CDW}=1$, $T_\eta/T_\mathrm{CDW}=0.99$, $b_\psi=b_\eta=0.25$, $c_\psi=c_\eta=0.5$.  For these parameters, the critical couplings are $d_2^*=2$, $d_4^*=0.47$, $d_4^{**}=0.35$ and $d_5^*=0.35$. The table reports the configuration of the two order parameters $\vec{\psi}$  and $\vec{\eta}$ with a hierarchy $a>b>c>0$. An arrow indicates the evolution of the order-parameter components upon lowering the temperature below the transition. We further specify whether FA or nematic order develops, and the corresponding symmetry subgroup. Note that $d_3$ mainly renormalizes the critical temperature of $\vec\eta$; on its own it leaves the $P\bar3$ and $P\bar1$ states degenerate. $^\dagger$For $d_4^{**}<|d_4|<d_4^*$, a further transition into $C2/m$ [$\vec\psi=(a,a,0)$, $\vec\eta=(a',-a',b')$] occurs at lower temperature, where the FA moment vanishes again. Finally, depending on the quadratic coefficients, the intermediate phase can be skipped, with a direct transition into the second SG. }
    \label{tab:di}
\end{table*}

\section{Numerical minimization of the free energy}
\label{app:numerics}

In Sec.~\ref{sec:GL}, the free energy was minimized numerically with respect to the six real components of $\vec\psi$ and $\vec\eta$ at fixed $b_\psi=b_\eta=0.25$, $c_\psi=c_\eta=0.5$, $a_\psi=1$ and $a_\eta=0.8$, $T_\mathrm{CDW}=1$ and $T_\eta=0.99$. 
We explored couplings of moderate strength, comparable to the single-field quartic coefficients ($b=0.25$, $c=0.5$), namely $|d_1|\le0.6$, $|d_2|\le0.6$, $|d_3|\le0.6$, $|d_4|\le0.6$ and $|d_5|\le0.6$, a range that includes the parameter sets shown in Fig.~\ref{fig:secIII} and within which the free energy is bounded from below for all parameter sets.

For each set of couplings, the global minimum was followed on a temperature grid $0.95\le T/T_\mathrm{CDW}\le1.002$ with step $5\times10^{-4}$. At each temperature, the minimum was taken as the lowest among local minimizations (quasi-Newton, as implemented in Mathematica's \texttt{FindMinimum}) started from three sets of initial conditions: (i) 20 random configurations with components uniformly distributed in $[-0.35,0.35]$; (ii) the minimum obtained at the previous temperature by continuation on cooling from the normal state; and (iii) the minimum obtained at the next lower temperature by continuation on heating from the lowest-temperature solution. Continuation seeds were perturbed by random displacements of order $10^{-4}$, so that the minimization does not remain at a saddle point when the previous state becomes unstable. As a benchmark, at 300 randomly chosen pairs of couplings and temperatures we compared the result with a global stochastic minimization (Mathematica's \texttt{NMinimize} with the \texttt{RandomSearch} method and the constraints $|\vec\psi|^2,|\vec\eta|^2\le1$). In all cases the two free energies differ by less than $5\times10^{-15}$ in relative terms.

To determine the order of the transitions, each transition between ordered phases was re-examined with cooling and heating continuations on a fine temperature grid with step $2\times10^{-6}$. The window was at most $\pm1.5\times10^{-3}$ around the transition, and never extended beyond half the distance to the neighbouring transitions of the same parameter set. A transition was classified as first order when two distinct local minima, with different order-parameter invariants ($|\vec\psi|$, $|\vec\eta|$, $\Phi_\mathrm{FA}$ and the moduli of the nematic vectors) and positive-definite Hessians, coexist over a finite temperature range, whose width defines the coexistence (hysteresis) interval. Otherwise, the transition was classified as continuous within this resolution, unless symmetry requires it to be first order, either because the two space groups are not related by a group--subgroup relation, or because the Landau expansion contains a cubic invariant of the critical order parameter. In the latter case, the coexistence interval is narrower than the resolution of the fine grid.

The strain–temperature phase diagrams of Figs.~\ref{fig:L1-L2-} and \ref{fig:exp} were obtained by minimizing $\mathcal{F}+\mathcal{F}_\varepsilon$ [Eq.~\eqref{eq:Feps}] on a regular grid with $-0.01225\le\varepsilon/\varepsilon_0\le0.01225$ in steps of $1.75\times10^{-4}$ and $0.98\le T/T_\mathrm{CDW}\le1.01$ in steps of $2.5\times10^{-4}$. At each grid point, the global minimum was obtained by stochastic global minimization (Mathematica's \texttt{NMinimize} with the \texttt{RandomSearch} method), which performs local minimizations from a set of random initial configurations and retains the lowest result.

\section{The nematic transition when $T_\mathrm{nem}<T_\mathrm{FA}$ is generically first order}
\label{app:Nem1st}

Out of the $C_3$-symmetric FA phase ($P\bar3$), a nematic transition is first order for one of two reasons: a direct transition into $C2/c$ or $C2/m$, where the FA moment vanishes, is discontinuous because these are not subgroups of $P\bar3$, while a transition into $P\bar1$, where the FA moment persists, admits a group–subgroup relation but is generically first order because of cubic invariants, as we now show.

We begin by considering nematic fluctuations around a $C_3$-symmetric FA state: 
\begin{equation}
\begin{split}
&\vec{\psi} = \frac{\psi_0}{\sqrt{3}}(1,1,1) + \delta_1(2,-1,-1) + \delta_2(0,\sqrt{3},-\sqrt{3}),\\
&\vec{\eta} = \frac{\eta_0}{\sqrt{3}}(1,1,1) + \epsilon_1(2,-1,-1) + \epsilon_2(0,\sqrt{3},-\sqrt{3}),
\end{split}
\end{equation}
where $(\delta_1,\delta_2)$ and $(\epsilon_1,\epsilon_2)$ parametrize fluctuations transverse to
$\frac{1}{\sqrt{3}}(1,1,1)$. The nematic vectors (defined in Eq.~\eqref{eq:nem_v}) read
\begin{equation}
\begin{split}
&\vec N_\psi = 2\sqrt{3}\,\psi_0\,(\delta_1,\delta_2) + 3\left(\delta_1^2-\delta_2^2,\,-2\delta_1\delta_2\right),\\
&\vec N_\eta = 2\sqrt{3}\,\eta_0\,(\epsilon_1,\epsilon_2) + 3\left(\epsilon_1^2-\epsilon_2^2,\,-2\epsilon_1\epsilon_2\right).
\end{split}
\end{equation}
It is now straightforward to verify that the quantities $|\vec{N}_\psi|^2$ and $|\vec{N}_\eta|^2$ appearing in the quartic free energy terms $\mathcal F_\psi$ and $\mathcal F_\eta$, see Eq. \eqref{eq:Fv_nem}, take the form
\begin{align}
|\vec N_\psi|^2 &= 12\,\psi_0^2(\delta_1^2+\delta_2^2) + 12\sqrt{3}\,\psi_0\left(\delta_1^3-3\delta_1\delta_2^2\right) \nonumber\\ &+ 9\,(\delta_1^2+\delta_2^2)^2,\\
|\vec N_\eta|^2 &= 12\,\eta_0^2(\epsilon_1^2+\epsilon_2^2) + 12\sqrt{3}\,\eta_0\left(\epsilon_1^3-3\epsilon_1\epsilon_2^2\right) \nonumber\\&+ 9\,(\epsilon_1^2+\epsilon_2^2)^2,
\end{align}
which contain the cubic terms $\delta_1^3-3\delta_1\delta_2^2$ and $\epsilon_1^3-3\epsilon_1\epsilon_2^2$
The presence of these cubic terms provides a mechanism for a first-order nematic transition within the $C_3$-symmetric FA phase. Indeed, cubic invariants are allowed by the symmetry of the two-component nematic fluctuations and generically render the transition first order at the mean-field level.

However, the coupling terms proportional to $d_i$ also contribute to the cubic expansion, and cancellations between the different contributions cannot be excluded on symmetry grounds alone. A continuous transition would require the effective cubic coefficient, projected onto the critical nematic mode, to vanish,
while the corresponding quartic coefficient remains positive. Such a cancellation would represent a fine-tuned critical point.

\bibliography{TiSe2}

\end{document}